\documentclass{article}

\usepackage{amsmath}
\usepackage{commath}
\usepackage{url}
\usepackage{cite}
\usepackage{color}
\usepackage{multirow}
\usepackage[margin=1.25in]{geometry}
\usepackage{graphicx}
\graphicspath{{./graphics/}}
\usepackage[font=small,margin=1cm]{caption}
\usepackage{pdfpages}

\newcommand\T{\rule{0pt}{2.6ex}}       % Top strut
\newcommand\B{\rule[-1.2ex]{0pt}{0pt}} % Bottom strut

\title{Studying Short-Range Correlations with Real Photon Beams at GlueX}

\author{
H.~Marukyan\\
A.~I.~Alikhanian National Science Laboratory\\
(Yerevan Physics Institute), 0036 Yerevan, Armenia\\
\and
M.~Patsyuk (Spokesperson) \\
Joint Institute for Nuclear Research, Dubna, Russia \\
\and
H.~Gao (Spokesperson)\\
Duke University, Durham, North Carolina 27708, USA\\
\and
M.~Kamel, L.~Guo\\
Florida International University, Miami, Florida 33199, USA\\
\and
D.~G.~Ireland, K.~Livingston, B.~Mckinnon, P.~Pauli, A.~Thiel  \\
University of Glasgow, Glasgow G12 8QQ, United Kingdom\\
\and
F.~Nerling\\
GSI Helmholtzzentrum f\"ur Schwerionenforschung GmbH,\\
D-64291 Darmstadt, Germany\\
\and
T.~Britton, M.~M.~Dalton, A.~Deur, H.~Egiyan, S.~Furletov, D.W.~Higinbotham,\\
 D.~Lawrence, D.~Mack, M.~McCaughan, L.~Pentchev, A.~Somov (Spokesperson), \\
H.~Szumila-Vance (Spokesperson), S.~Taylor, B.~Zihlmann\\
Thomas Jefferson National Accelerator Facility,\\ Newport News, Virginia 23606, USA\\
\and
A.~Ashkenazi, R.~Cruz-Torres, A.~Denniston, C.~Fanelli, 
O.~Hen (Spokesperson\footnote{Contact Person: \texttt{hen@mit.edu}}),\\
D.~Nguyen, A.~Papadopoulou, J.~R.~Pybus, E.P.~Segarra\\
Massachusetts Institute of Technology, Cambridge, Massachusetts 02139, USA\\
\and
D.~Dutta (Spokesperson)\\
Mississippi State University, Mississippi State, Mississippi 29762, USA\\
\and
V.~Berdnikov, D.~Romanov, S.~Somov\\
National Research Nuclear University Moscow Engineering Physics Institute, \\Moscow 115409, Russia\\
\and
C.~Salgado\\
Norfolk State University, Norfolk, Virginia 23504, USA\\
\and
R.~Pedroni\\
North Carolina A\&T State University, Greensboro, North Carolina 27411, USA\\
\and
T.~Black, L.~Gan\\
University of North Carolina at Wilmington,\\
Wilmington, North Carolina 28403, USA\\
\and
A.~Beck, I.~Korover, and S.~Maytal-Beck\\
Nuclear Research Center Negev, Beer-Sheva 84190, Israel\\
\and
M.~Amaryan, F.~Hauenstein, L.B.~Weinstein (Spokesperson)\\
Old Dominion University, Norfolk, Virginia 23529, USA\\
\and
W.~Brooks, H.~Hakobyan, S.~Kuleshov, C.~Romero\\
Universidad T\'ecnica Federico Santa Mar\'ia, Casilla 110-V Valpara\'iso, Chile\\
\and
B.~Schmookler\\
State University of New York at Stony Brook, New York, 11794\\
\and
G.~Johansson, E.~Piasetzky (Spokesperson)\\
Tel-Aviv University, Tel Aviv 69978, Israel\\
\and
W.~J.~Briscoe, S.~Fegan, S.~Ratliff, A.~Schmidt (Spokesperson), E.~Seroka, \\P.~Sharp, I.~I.~Strakovsky\\
The George Washington University, Washington, D.C. 20052, USA\\
}

\date{\vspace{-5ex}}

\begin{document}

\maketitle

\vspace{10pt}
\noindent\underline{Theory Support}\\
\begin{tabular}{l p{4.5in}}
M. Strikman & Pennsylvania State University, State College, PA.\\
G.A. Miller & University of Washington, Seattle, WA. \\
A.B. Larionov & Frankfurt Institute for Advanced Studies (FIAS),
and National Research Centre ``Kurchatov Institute'', Moscow, Russia.\\
M. Sargsian & Florida International University, Miami, FL. \\
L. Frankfurt & Tel-Aviv University, Tel Aviv, Israel.
\end{tabular}

\begin{abstract}

The past few years has seen tremendous progress in our understanding of short-range
correlated (SRC) pairing of nucleons within nuclei, much of it coming from electron
scattering experiments leading to the break-up of an SRC pair. The interpretation
of these experiments rests on assumptions about the mechanism of the reaction. 
These assumptions can be directly tested by studying SRC pairs using alternate probes,
such as real photons. We propose a 30-day experiment using the Hall D photon beam, 
nuclear targets, and the GlueX detector in its standard configuration 
to study short-range correlations with photon-induced
reactions. Several different reaction channels are possible, and we project sensitivity
in most channels to equal or exceed the 6~GeV-era SRC experiments from Halls A and B. 
The proposed experiment will therefore decisively test the phenomena of $np$ dominance,
the short-distance $NN$ interaction, and reaction theory, while also providing new
insight into bound nucleon structure and the onset of color transparency.
\end{abstract}

\begin{center}
\includegraphics[width=0.8\textwidth]{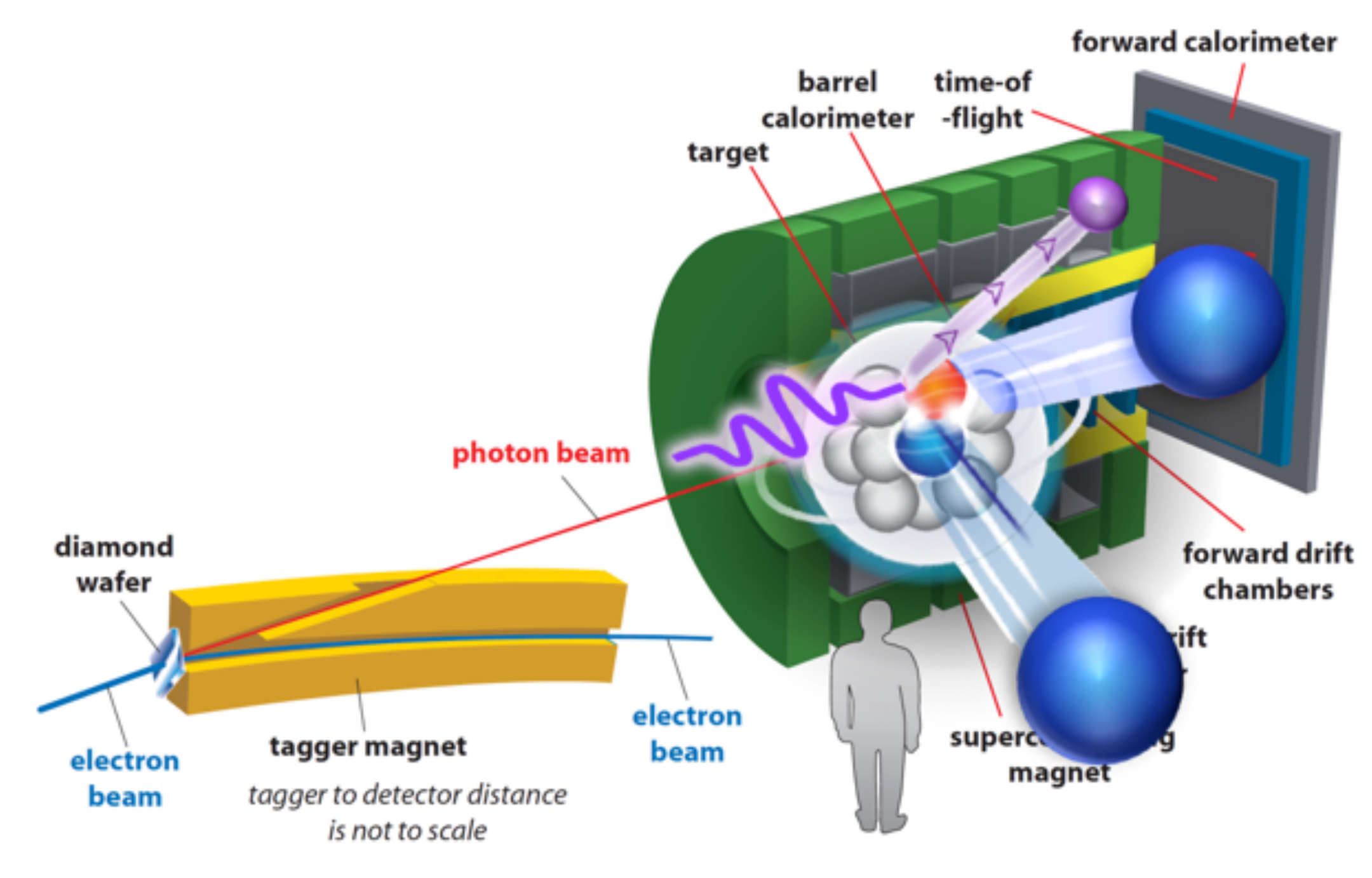}
\end{center}

\section{Introduction}
Since the 1950s much effort has been devoted to understanding the detailed characteristics and origin
of the nucleon-nucleon interaction, and how it forms atomic nuclei. The development of modern
superconducting accelerators---with high energy, high intensity and high duty factor---has enabled
scattering experiments that resolve the structure and dynamics of both individual nucleons and nucleon
pairs in nuclei, allowing significant progress. While many breakthroughs have been made, much still
remains to be understood.

Early measurements of inclusive $(e,e')$ and single proton knockout $(e,e'p)$ processes helped
establish the shell-structure of nuclei and probe properties of nucleons bound in different nuclear
shells~\cite{Hofstadter:1956qs,DeForest:1966ycn,Kelly:1996hd}. While lending credence to many shell
model predictions, these measurements also revealed that in nuclei ranging from lithium to lead,
proton-knockout cross-sections are only 60\%--70\% of the mean-field one-body-based theoretical
expectation~\cite{Lapikas:1003zz,Dickhoff:2004xx}, highlighting the need to consider higher order
two-body effects that go beyond the traditional mean-field approximation.

Electron-scattering experiments, analyzed within a high-resolution theoretical framework, suggest
that about 20\% of the nucleons in nuclei are part of strongly interacting close-proximity nucleon
pairs, with large relative ($k_{rel} > k_F \approx 250$ MeV/c) and small center-of-mass (C.M.)
momenta ($k_{CM} < k_F$). These are referred to as short-range correlated (SRC)
pairs~\cite{Hen:2016kwk,Atti:2015eda,Arrington:2011xs,Frankfurt:2008zv}. Nucleons that are part of
SRC pairs are absent in the one-body shell-model description of the data. Their formation can
therefore explain some of the discrepancy between the measured and calculated single-proton knockout
cross-sections~\cite{Dickhoff:2004xx,Paschalis:2018zkx} and have wide-spread implications for
different phenomena in nuclear-, particle- and astro-physics. 

A large part of our understanding of SRC pairs comes from measurements
of exclusive two-nucleon knockout reactions $(e,e'NN)$. In these
experiments, a high-energy electron scatters off the nucleus, leading
to the knockout of a nucleon with large missing momentum that is
balanced by the emission of a single recoil nucleon, leaving the
residual $A-2$ system with both low momentum and low excitation energy.

The very successful JLab SRC program resulted in multiple experimental
papers published in Nature~\cite{Schmookler:2019nvf,Duer:2018sby}, 
Science~\cite{Hen:2014nza,Subedi:2008zz},
PRL~\cite{Shneor:2007tu,Korover:2014dma,Duer:2018sxh,Cohen:2018gzh},
Physics Letters B~\cite{Hen:2012yva} and more, with new results currently
undergoing peer-review~\cite{Duer:2018sjb,Cruz-Torres:2019bqw}.  These results showed that almost all
high-momentum nucleons in nuclei belong to SRC
pairs~\cite{Korover:2014dma,Duer:2018sxh,Piasetzky:2006ai,Subedi:2008zz,Shneor:2007tu,Tang:2002ww}.
For relative momenta between 300--600~MeV$/c$ (just above $k_F$),
neutron-proton ($np$) SRC pairs predominate over proton-proton ($pp$) and
neutron-neutron ($nn$) pairs by a factor of about 20, in both light and
heavy nuclei~\cite{Hen:2014nza,Korover:2014dma,Duer:2018sxh}.  This phenomenon, commonly referred to as
``$np$-dominance,''~\cite{Duer:2018sby,Hen:2014nza,Korover:2014dma,Duer:2018sxh,Piasetzky:2006ai,Subedi:2008zz,Shneor:2007tu,Tang:2002ww},
is driven by the tensor nature of the $NN$ interaction in the quoted momentum range~\cite{Schiavilla:2006xx,Sargsian:2005ru,Alvioli:2007zz} and
indicates that increasing the fraction of neutrons in a nucleus
increases the fraction of protons that are part of SRC pairs~\cite{Duer:2018sby,Sargsian:2012sm,Ryckebusch:2018rct}.

These experimental findings inspired a broad complementary program of
theoretical and phenomenological studies of SRCs and their impact on
various nuclear phenomena, including the internal structure of
nucleons bound in nuclei~\cite{Weinstein:2010rt,Hen:2012fm,Hen:2013oha,cern:courier,Schmookler:2019nvf}, neutrinoless double beta decay
matrix elements~\cite{Cruz-Torres:2017sjy,Kortelainen:2007rn,Kortelainen:2007rh,Kortelainen:2007mn,Menendez:2008jp,Simkovic:2009pp,Engel:2009gb},
nuclear charge radii~\cite{Weiss:2018zrd}, and the nuclear
symmetry energy and neutron star properties~\cite{Frankfurt:2008zv,Hen:2014yfa,Cai:2017rwj}.

However, almost all of our understand of SRC comes from electron
scattering, with only a single proton scattering C$(p,ppn)$ measurement~\cite{Tang:2002ww}.
Thus, the interpretation of these experimental results relies
on an assumed electron interaction mechanism at large momentum
transfers (detailed in section~\ref{sec:progress} below). Different assumptions could
lead to different interpretations. It is crucial to study SRCs with
different, non-electron, probes, in order to validate the reaction
mechanism assumptions and the connection between the experimental
results and their interpretation in terms of SRC pairs.

This proposal describes an experiment to study SRCs using photo-production
reactions with real photons on nuclear targets to determine the probe-dependence of SRC 
measurements. The new data will complement the above-mentioned electron 
scattering studies and yield stringent constraints on possible reaction
mechanisms that could complicate the interpretation of the data.  
Photo-nuclear reactions have significantly different sensitivity to
meson exchange currents, are dominated by the transverse response function
with backward emitted recoil nucleons (as oppose to both longitudinal and
transverse response functions and forward emitted recoil nucleons in $x_B>1$
electron scattering kinematics). If the reaction mechanisms at these high
momentum transfer reactions ($Q^2$, $|t|$, $|u| > 2$~GeV$^2$) are
indeed understood to the expected level, we should be able to confirm
the observed neutron-proton pair dominance and the $A$-dependence of
SRC-pair abundances through this experiment. We note that nuclear targets
have been considered as part of the future plan for the GlueX Detector,
as laid out by the recent GlueX collaboration white paper~\cite{gluex:whitepaper}. 

We propose a 30-day measurement using the real photon beam in Hall D, 
three nuclear targets ($d$, $^4$He, and $^{12}$C), and the GlueX detector
in its standard configuration. The main goal of the experiment will be
to study short-range correlations using photon-induced reactions. The
experiment can additionally provide information on bound nucleon structure
(discussed in section~\ref{sec:bns}) and color-transparency
(discussed in section~\ref{sec:ct}). 

\section{Recent progress in the quantitative study of SRCs}
\label{sec:progress}

The study of short-range correlations is a broad subject. It covers a large body of 
experimental and theoretical work, as well as phenomenological studies of the implications
of SRCs for various phenomena in nuclear, particle and astro-physics. The discussion below
is focused primarily on recent experimental activities co-led by the spokespersons, and theoretical
developments that are most relevant for the objectives of the current proposal. A full 
discussion of SRC physics is available in a recent RMP review~\cite{Hen:2016kwk}, as well
as in a theory-oriented review~\cite{Atti:2015eda}.

\begin{figure}[htpb]
\centering
\includegraphics{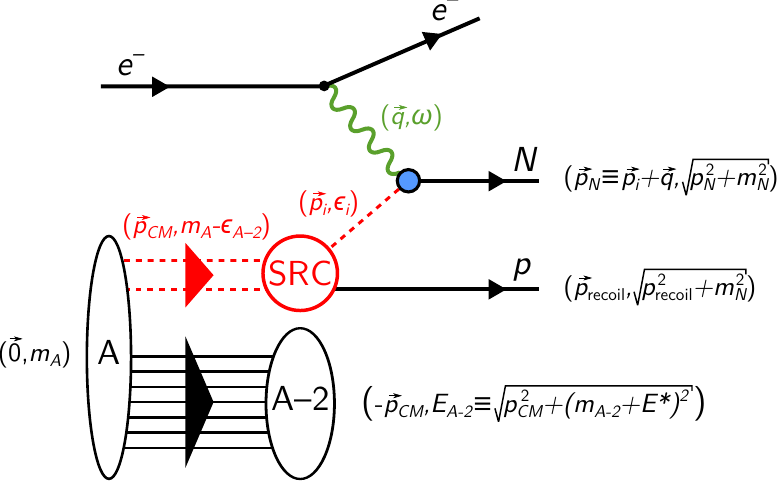}
\caption{\label{fig:reaction_e}
Diagrammatic representation and kinematics of the triple-coincidence $A(e,e'Np)$ reaction within 
the SRC breakup model. Dashed red lines represent off-shell particles. Open ovals represent un-detected
systems. Solid black lines represent detected particles. The momentum and energy of the particles are also indicated.}
\end{figure}

Previous studies of SRCs have used measurements of Quasi-Elastic (QE) electron scattering 
at large momentum-transfer, see Fig.~\ref{fig:reaction_e}. Within the single-photon exchange approximation, 
electrons scatter from the nucleus by transferring a virtual photon carrying momentum $\vec{q}$
and energy $\omega$. In the one-body view of QE scattering, the virtual photon is absorbed by 
a single off-shell nucleon with initial energy $\epsilon_i$ and momentum $\vec{p}_i$. If the 
nucleon does not re-interact as it leaves the nucleus, it will emerge with momentum $\vec{p}_N = \vec{p}_i + \vec{q}$
and energy $E_N=\sqrt{p_N^2 + m_N^2}$. Thus, we can approximate the initial momentum and energy
of that nucleon using the measured missing momentum, $\vec{p}_i \approx \vec{p}_\text{miss} \equiv \vec{p}_N - \vec{q}$,
and missing energy, $\epsilon_i \approx m_N - \epsilon_\text{miss} \equiv \epsilon_N - \omega$.
When $\vec{p}_\text{miss} > k_F$, the knockout nucleon is expected to be part of an SRC pair
\cite{Hen:2016kwk,Atti:2015eda,Arrington:2011xs,Frankfurt:2008zv,Korover:2014dma,Piasetzky:2006ai,Subedi:2008zz}. 
The knockout of one nucleon from the pair should therefore be accompanied by the 
simultaneous emission of the second (recoil) nucleon with momentum $\vec{p}_\text{recoil} \approx -\vec{p}_\text{miss}$.
At the relevant high-$Q^2$ of our measurements ($>1.7$--$2.0$~GeV$/c$), the differential $A(e,e'p)$ 
cross-sections can be approximately factorized as~\cite{Kelly:1996hd,DeForest:1983ahx}:
\begin{equation}
\frac{d^6\sigma}{d\Omega_{k'} d\epsilon_{k'} d\Omega_N d\epsilon_N} = p_N \epsilon_N \cdot \sigma_{ep} \cdot \mathcal{S}(p_i,\epsilon_i),
\label{eq:spectral}
\end{equation}
where $k' = (k', \epsilon_{k'})$ is the final electron four-momentum, $\sigma_{ep}$ is the off-shell 
electron-nucleon cross-section~\cite{DeForest:1983ahx}, and $\mathcal{S}(p_i,\epsilon_i)$ is the nuclear
spectral function that defines the probability for finding a nucleon in the nucleus with momentum $p_i$ 
and energy $\epsilon_i$. Different models of the $NN$ interaction can produce different spectral functions
that lead to different cross-sections. Therefore, exclusive nucleon knockout cross-sections analyzed with this method
are sensitive to the $NN$ interaction.

In the case of two-nucleon knockout reactions, the cross-section can be factorized in a similar manner 
to Eq.~\ref{eq:spectral} by replacing the single-nucleon spectral function with the two-nucleon decay
function $\mathcal{D}_A(p_i,p_\text{recoil},\epsilon_\text{recoil})$~\cite{Frankfurt:2008zv,Piasetzky:2006ai,Frankfurt:1988nt}.
The latter represents the probability for a hard knockout of a nucleon with initial momentum $\vec{p}_i$, 
followed by the emission of a recoil nucleon with momentum $\vec{p}_\text{recoil}$. $\epsilon_\text{recoil}$ 
is the energy of the $A-1$ system, composed of the recoil nucleon and residual $A-2$ nucleus.  

Non-QE reaction mechanisms that add coherently to the measured cross-section can lead to high-$p_\text{miss}$ 
final states that are not due to the knockout of nucleons from SRC pairs, thus breaking the factorization shown
in Eq.~\ref{eq:spectral}. To address this, the measurements discussed here are carried out at anti-parallel 
kinematics with $p_\text{miss}\ge 300$~MeV$/c$, $Q^2\equiv q^2 - \omega^2 \ge 1.7$~(GeV$/c$)$^2$, and 
$x_B \equiv Q^2 / 2 m_N \omega \ge 1.2$, where such non-QE reaction mechanisms were shown to be
suppressed~\cite{Hen:2016kwk,Atti:2015eda,Arrington:2011xs,Frankfurt:2008zv,Colle:2015lyl,Sargsian:2001ax}.

For completeness, we note that from a theoretical standpoint, the reaction diagram shown in Fig.~\ref{fig:reaction_e}
can be viewed as a `high-resolution' starting point for a unitary-transformed calculation~\cite{More:2017syr}. 
Such calculations would soften the input $NN$ interactions and turn the electron scattering operators
from one-body to many-body. This `unitary-freedom' does not impact cross-section calculations but does
make the extracted properties of the nuclear ground-state wave-function (e.g. the spectral function)
depend on the assumed interaction operator. This discussion focuses on the high-resolution electron
interaction model of Fig.~\ref{fig:reaction_e}, as it constitutes the simplest reaction picture that is consistent
with both the measured observables~\cite{Hen:2016kwk,Atti:2015eda,Arrington:2011xs,Frankfurt:2008zv} 
and various reaction and ground-state ab-initio calculations~\cite{Carlson:2014vla}.

\subsection{Short-Distance $NN$ Interaction at the Generalized Contact Formalism}
\label{ssec:gcf}
Precision SRC studies are only feasible if one has the ability to quantitatively relate
experimental observables to theoretical calculations, ideally ones starting from the 
fundamental $NN$ interaction and accounting for all relevant reaction mechanisms. This
is a challenging endeavor, as un-factorized ab-initio calculations of high-$Q^2$ nucleon
knockout cross-sections are currently unfeasible for $A>3$ nuclei. Even the simple 
factorized approximation of Eq.~\ref{eq:spectral} requires knowledge of the nuclear
spectral function that, at the moment, cannot be calculated using ab-initio techniques
for high-momentum states in finite nuclei~\cite{Carlson:2014vla}. 

To help overcome this challenge, the effective
two-body Generalized Contact Formalism (GCF) was recently developed,~\cite{Weiss:2016obx,Weiss:2018tbu,Weiss:2015mba}
that allows calculating factorized cross-sections, within a scale-separated approximation, 
using the underlying $NN$ interaction as input~\cite{Duer:2018sxh,Weiss:2018tbu}. This is 
done by providing a factorized model of the short-distance / high-momentum part of the many-body
nuclear wave function leveraging the separation between the energy scales of the $A-2$ system
(low energy) and the SRC pair (medium energy). Considering a high-$Q^2$ scattering reactions 
such as in Fig.~\ref{fig:reaction_e} adds a third energy scale of the virtual photon (high-energy)
that justifies the factorized approximation of Eq.~\ref{eq:spectral}.

The GCF provides a consistent model for nuclear two-body momentum distribution at high-momenta 
and at short-distance, as well as for two-body continuum states of the nuclear spectral and decay functions.  
Recent studies of the GCF:
\begin{itemize}
\item Demonstrated its ability to reproduce many-body ab-initio calculated nucleon momentum
distributions in nuclei from $^4$He to $^{40}$Ca, above $k_F$, to $\approx 10\%$ accuracy\cite{Weiss:2016obx};
\item Extracted consistent SRC abundances (i.e., nuclear contacts) from ab-initio calculations of 
two-nucleon distributions in both coordinate and momentum space and from experimental data~\cite{Weiss:2016obx}; and 
\item Derived a new factorized expression for the nuclear correlation function with implications
for calculations of double beta decay matrix elements~\cite{Cruz-Torres:2017sjy} and demonstrated
its relation to single-nucleon charge distribution measurements \cite{Weiss:2018zrd}.
\end{itemize}
%% *** IF we want to append scale and scheme paper ***
%%The GCF has numerous other applications currently being explored, and has offered new insight into 
%%scale separation in nuclear structure. As an example, a draft of a paper in preparation by the spokespersons
%%and collaborators exploring the scale- and scheme-dependence of nuclear contacts is appended to this proposal.

The main application of the GCF germane to this proposal is the modeling of the nuclear
spectral and decay functions~\cite{Weiss:2018tbu}, allowing calculations of nucleon knockout cross-sections. 
For example, using Eq.~\ref{eq:spectral} and the reaction model of Fig.~\ref{fig:reaction_e}, the $A(e,e'NN)$
cross-section can be expressed within the GCF as~\cite{Duer:2018sxh}:
\begin{equation}
\label{eq:gcf_cs}
\frac{d^8\sigma}{dQ^2 dx_B d\phi_k d^3 \vec{p}_{CM} d\Omega_\text{recoil}}
= K \cdot \sigma_{eN} \cdot n(\vec{p}_{CM})\cdot
\left[ \sum\limits_\alpha C_\alpha \cdot |\tilde{\varphi}^\alpha(|\vec{p}_{CM} - 2\vec{p}_\text{recoil}|)|^2\right],
\end{equation}
where subscripts `$N$' and `recoil' stand for the leading and recoil nucleon respectively, $K$ is a kinematic term,
(detailed in Ref.~\cite{Duer:2018sxh}), $\sigma_{eN}$ is the off-shell electron-nucleon cross-section, and 
$\alpha$ represents the spin and isospin quantum numbers of SRC pairs. $\tilde{\varphi}^\alpha$, $n(\vec{p}_{CM})$, and $C_\alpha$
respectively describe the relative motion, CM motion, and abundances of SRC pairs with quantum numbers $\alpha$.
The functions $\tilde{\varphi}^\alpha$ are universal SRC pair relative momentum distributions, obtained by solving
the zero-energy two-body Schr\"{o}dinger equation of an $NN$ pair in quantum state $\alpha$ using an input $NN$ potential model.
$n(\vec{p}_{CM})$ is the SRC pair CM momentum distribution, given by a three-dimensional Gaussian with width of
$150\pm 20$~MeV$/c$~\cite{Cohen:2018gzh,CiofidegliAtti:1995qe,Colle:2013nna}. $C_\alpha$ are the nuclear contact
terms that determine the relative abundance of SRC pairs in quantum state $\alpha$. These are obtained through
the analysis of ab-initio many-body calculations of two-nucleon densities~\cite{Weiss:2016obx,Weiss:2015mba,Wiringa:2013ala}.

\begin{figure}[htpb]
\centering
\includegraphics[height=8.5cm]{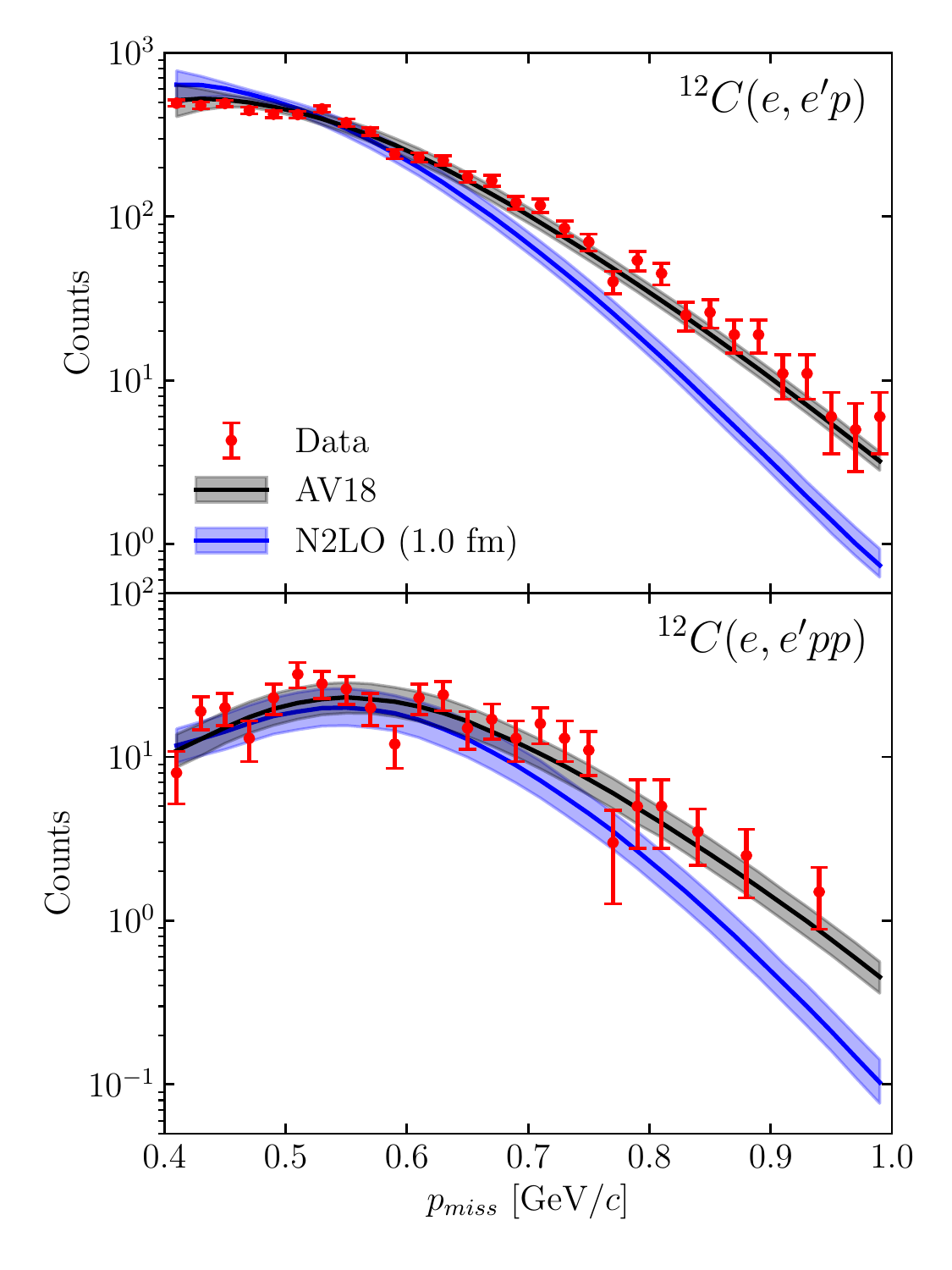}
\includegraphics[height=8.5cm]{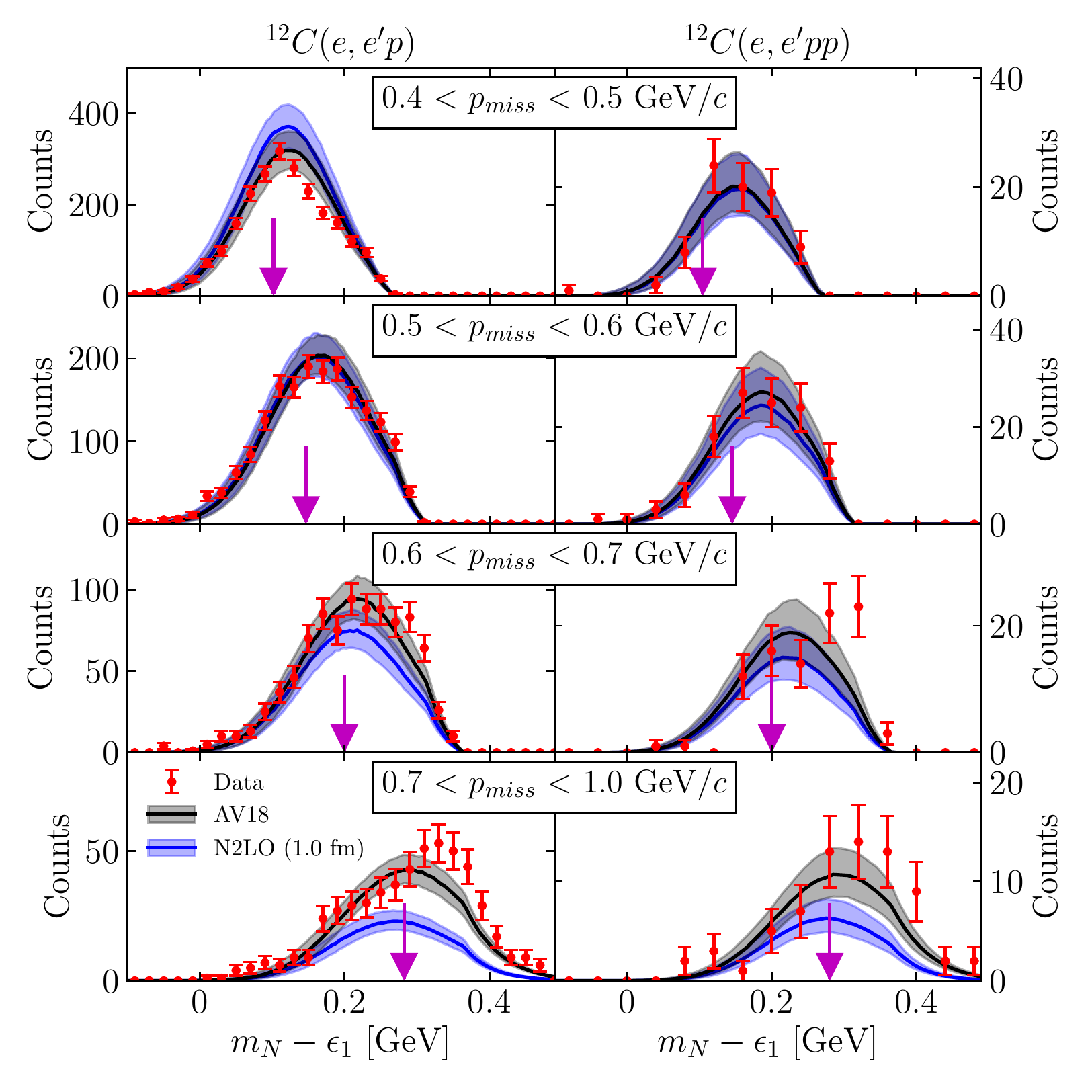}
\caption{\label{fig:gen_plots} Left panel: the $p_\text{miss}$ dependence of the $^{12}$C$(e,e'p)$ (top) 
and $^{12}$C$(e,e'pp)$ (bottom) event yields. Points show the measured data. Bands 
show the GCF calculations using the N2LO(1.0fm) (blue) and AV18 (black) interactions. 
Right panel: the $\epsilon_\text{miss}$ dependence of the $^{12}$C$(e,e'p)$ (left column)
and $^{12}$C$(e,e'pp)$ (right column) event yields in four different ranges of $p_\text{miss}$. 
The purple arrow indicates the expected $\epsilon_\text{miss}$ for standing SRC pair breakup 
with a missing-momentum that is equal to the mean value of the data.}
\end{figure}

\begin{figure}[htpb]
\centering
\includegraphics[width=\textwidth]{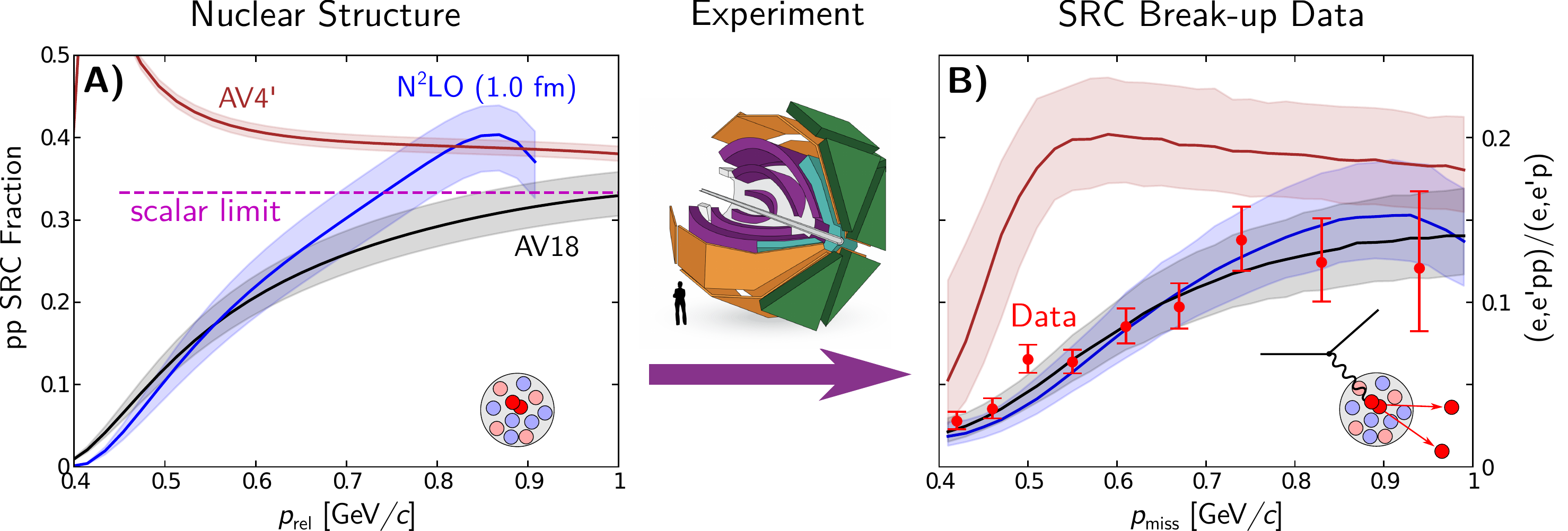}
\caption{\label{fig:pp2p} A: the $pp$ pair fraction in $^{12}$C as predicted by GCF using AV18,
AV4', and Chiral N2LO(1.0 fm) interactions. B: the ratio of $^{12}$C$(e,e'pp)$ to $^{12}$C$(e,e'p)$
event yields for data (red points) and GCF (bands), including all experimental effects. Both the
AV18 and N2LO(1.0 fm) interactions are consistent with data, and show an increase from a tensor-dominated
regime at $p_\text{miss}=0.4$~GeV$/c$ to scalar spin-independent regime approaching $p_\text{miss}=1$~GeV$/c$.
The AV4' interaction, which has no tensor component, leads to predictions that are inconsistent with data.
}
\end{figure}

Ref.~\cite{Duer:2018sxh} shows the first comparisons between the prediction of Eq.~\ref{eq:gcf_cs} and measured $A(e,e'Np)$
 cross-section ratios. In Figs.~\ref{fig:gen_plots} and \ref{fig:pp2p} we showcase our most recent results,
where Eq.~\ref{eq:gcf_cs} is used to calculate the individual $(e,e'p)$ and $(e,e'pp)$ cross-sections in the 
kinematics of our SRC measurements. The calculation was done using two $NN$ 
interaction models to obtain $\tilde{\varphi}^\alpha$: the phenomenological AV18~\cite{Wiringa:1994wb}, and Chiral EFT-based
local N2LO(1.0 fm)~\cite{Gezerlis:2013ipa}. Nuclear contacts $C_\alpha$ and width of the CM momentum distribution were obtained
from theoretical calculations~\cite{Weiss:2016obx,Weiss:2015mba,CiofidegliAtti:1995qe,Colle:2013nna,Wiringa:2013ala} and
nuclear transparency and single-charge exchange reaction effects were accounted for as detailed in the online supplementary
materials of Ref.~\cite{Duer:2018sxh}, using the calculations of Ref.~\cite{Colle:2015lyl}. The model systematic uncertainty
is determined from the uncertainties in the GCF input parameters and reaction effects correction factors.

The left panel of Fig.~\ref{fig:gen_plots} shows the $p_\text{miss}$ dependence of the measured
and GCF-calculated $^{12}$C$(e,e'pp)$ and $^{12}$C$(e,e'p)$ event yields for the two interactions.
The AV18 interaction is observed to describe both $(e,e'p)$ and $(e,e'pp)$ data over the entire
measured $p_\text{miss}$ range. The N2LO(1.0 fm) interaction agrees with the data up to its cutoff 
and, as expected, decreases exponentially above it. 

The right panel of Fig.~\ref{fig:gen_plots} shows the $\epsilon_\text{miss}$-$p_\text{miss}$ correlation
for the $^{12}$C$(e,e'pp)$ and $^{12}$C$(e,e'p)$ reactions. The average value of $m_N - \epsilon_1$ is 
observed to increase with $p_\text{miss}$, peaking at the expected value for the breakup of a standing
SRC pair (indicated by the purple arrows) for both reactions. The GCF calculations follow the same trend.
However, the AV18 interaction agrees with the data over the entire $\epsilon_\text{miss}$-$p_\text{miss}$
range, while the chiral interactions under predict at the highest $p_\text{miss}$.

Fig.~\ref{fig:pp2p} considers the $^{12}$C$(e,e'pp)/^{12}$C$(e,e'p)$ yield ratio, a measure of the
impact of the tensor force in the $NN$ interaction. In this figure, the AV18 and the chiral N2LO(1.0~fm)
interactions are compared to the AV$4'$ interaction, which does not include a tensor force. The right panel
shows the data yield ratio as well as the GCF-calculated yield ratio. Both the data, and the calculations with
the AV18 and N2LO(1.0~fm) interactions show the $pp$ fraction increasing
with $p_\text{miss}$, consistent with a transition from tensor- to scalar-dominated regions of the
interaction~\cite{Korover:2014dma}. By contrast, the calculation with the AV$4'$ interaction over-predicts
the fraction of $pp$ pairs observed in the data. 

The left panel shows the fraction of $pp$ pairs in $^{12}$C as predicted by the GCF formalism as a 
function of ${p}_\text{rel.} \equiv \frac{1}{2}|\vec{p}_\text{miss} - \vec{p}_\text{recoil}|$. 
The AV18 and N2LO(1.0~fm) interactions approach limit predicted by a purely spin-independent
interaction. The AV$4'$ interaction, without a tensor force, predicts a $pp$ fraction above this
scalar limit.

We note that our confidence in these results is also supported by the fact that the GCF-based
calculations describe well numerous other measured kinematical distributions not shown here due to a lack of space.
Thus, the results presented here showcase the use of high-$Q^2$ electron scattering data to quantitatively study
the nuclear interaction at very large momenta. 

It is interesting to note that for the AV18 interaction, we observe good agreement with the data up to 1~GeV$/c$,
which corresponds to SRC configurations with nucleons separated by a distance smaller than their radii~\cite{Neff:2016ajx}. 
As discussed below, previous studies indicated that in such extreme conditions the internal quark-gluon structure
of SRC nucleons can well be modified as compared with that of free 
nucleons~\cite{Hen:2016kwk,Schmookler:2019nvf,Chen:2016bde,CiofidegliAtti:2007ork,Kulagin:2010gd}.
The ability of the AV18-based GCF calculation to reproduce our data over the entire measured $\epsilon_\text{miss}$-$p_\text{miss}$
range suggests that such modifications do not significantly impact the effective modeling of the nuclear interaction, offering support
for using point-like nucleons as effective degrees of freedom for modeling of nuclear systems up to very high densities.

\subsection{Two-Nucleon Knockout Reactions}
\label{ssec:NN}

The above-mentioned results constitute some of the most advanced, ongoing (i.e. unpublished), 
analysis that utilizes the scale-separated GCF to calculate factorized nucleon-knockout
cross-sections using different models of the $NN$ interaction. These studies are made 
possible by the vast progress made in the study of SRCs using hard knockout reactions over
the last decade. 
Below, we review key published results from initial measurements of nuclei from $^4$He to $^{208}$Pb.
%The PI's contribution to this experimental program resulted in papers published in Nature~\cite{Schmookler:2019nvf,Duer:2018sby}, Science~\cite{Hen:2014nza}, PRL~\cite{Korover:2014dma,Duer:2018sxh,Cohen:2018gzh}, and Physics Letters~\cite{Hen:2012yva}, as well as others recently submitted for publication~\cite{Duer:2018sjb}.

\subsubsection{$np$-SRC dominance and the tensor interaction}

First measurements of exclusive SRC pair breakup reactions focused primarily on probing the 
isospin structure of SRC pairs. These experiments were initially done at BNL using hadronic
(proton) probes on $^{12}$C, and continued at JLab with leptonic (electron) probes on $^4$He,
$^{12}$C, $^{27}$Al, $^{56}$Fe and $^{208}$Pb. 

Focusing on a missing momentum range of 300--600~MeV$/c$, comparisons of the measured $A(e,e'p)$
and $A(e,e'pN)$ cross-section indicated that the full single-proton knockout cross-section is
exhausted by the two-nucleon knockout cross-sections, i.e., the data were consistent with every
$(e,e'p)$ event having the correlated emission of a recoil nucleon~\cite{Korover:2014dma,Duer:2018sxh,Piasetzky:2006ai,Subedi:2008zz}.
A common interpretation of these results is that the nucleon momentum distribution above $k_F$ is
dominated by nucleons that are members of SRC pairs.

\begin{figure}[htpb]
\centering
\includegraphics[width=0.6\textwidth]{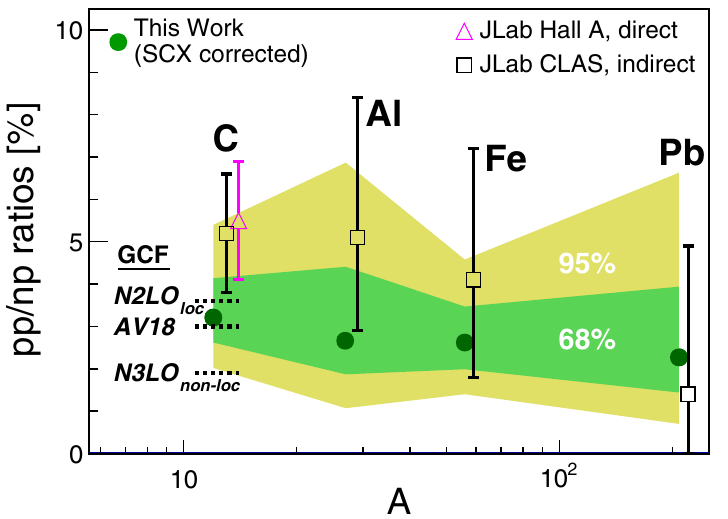}
\caption{\label{fig:meytal_pp_np}
$np$-SRC dominance in nuclei from $^{12}$C to $^{208}$Pb extracted from $A(e,e'Np)$ and $A(e,e'p)$ 
measurements~\cite{Hen:2014nza,Duer:2018sxh,Subedi:2008zz}, compared with GCF calculations~\cite{Duer:2018sxh}.
}
\end{figure}

\begin{figure}[htpb]
\centering
\includegraphics[height=5cm]{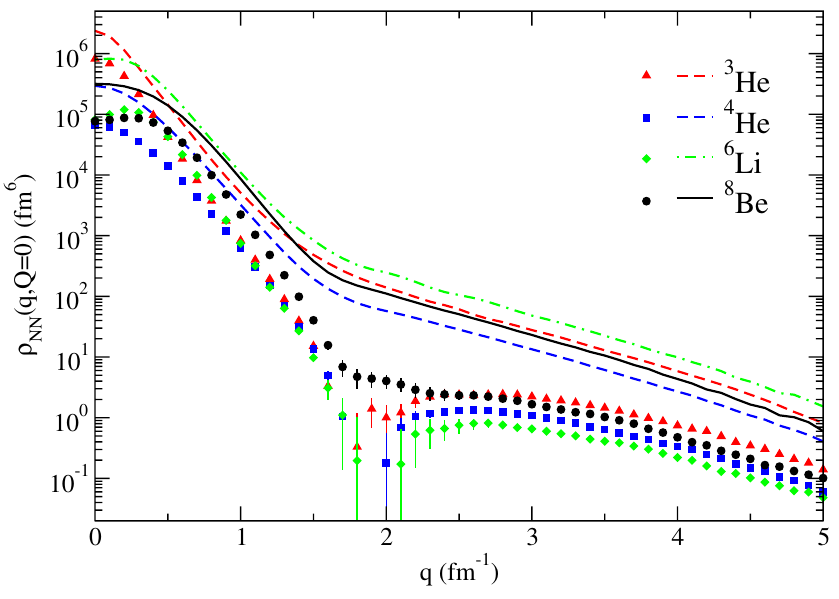}
\includegraphics[height=5cm]{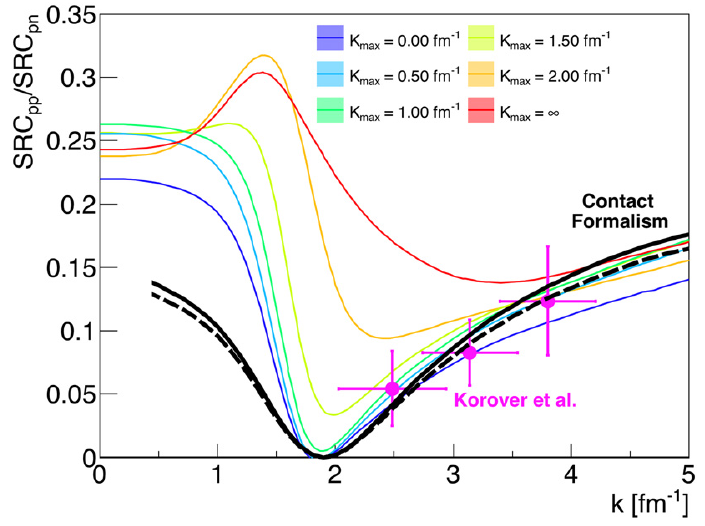}
\caption{\label{fig:np_dom} Left: calculated $pp$ (points) and $np$ (lines) stationary pair
momentum densities in light nuclei~\cite{Schiavilla:2006xx}. Right: measured and calculated
$^4$He $pp/np$ pair density ratios as a function of the pair relative momentum~\cite{Weiss:2016obx}.
}
\end{figure}

Furthermore, the measured $A(e,e'pn)$ and $A(e,e'np)$ cross-sections were found to be significantly
higher than the $A(e,e'pp)$ cross-section. This finding, consistently observed in all measured nuclei,
was interpreted as evidence for $np$-SRC pairs being about 20$\times$ more abundant than $pp$-SRC pairs
(Fig.~\ref{fig:meytal_pp_np}). From a theoretical standpoint, this $np$-SRC predominance was interpreted
as resulting from the dominance of the tensor part of the $NN$ interaction at the probed sub-fm
distances~\cite{Hen:2016kwk,Atti:2015eda,Schiavilla:2006xx,Sargsian:2005ru,Alvioli:2007zz} (see
Fig.~\ref{fig:np_dom}.

It should be pointed out that, on average, the tensor part of the $NN$ interaction is long-ranged
and small compared to the dominant scalar part. However, studies of the deuteron suggest that its
second order effect, viewed as a two-pion exchange term, becomes important in the momentum range
where the scalar force approaches zero ($\approx 0.75$--1~fm)\cite{Hen:2016kwk}. At shorter 
distances, i.e., higher relative momenta, the dominance of the tensor interaction is expected to be
washed out, which would manifest in an increase in the fraction of $pp$-SRC pairs with much larger
missing momentum. Fig.~\ref{fig:np_dom} shows the measured increase in the fraction of $pp$-SRC pairs~\cite{Korover:2014dma},
which is overall consistent with theoretical expectation based on calculations of two-nucleon momentum
distributions~\cite{Wiringa:2013ala} and their GCF representation~\cite{Weiss:2016obx}.
The large error bars of the $^4$He data makes it hard to draw any conclusive quantitative conclusions
on the $NN$ interaction beyond the tensor limit. However, as shown above, the combination of improved
data, and recent theoretical developments (such as the GCF), allows addressing such issues.

\subsubsection{SRC pair C.M. motion}

Measurements of exclusive two-nucleon knockout reactions allow us to probe the detailed characteristics of SRC pairs,
going beyond their isospin structure. One such property of interest is the C.M.~motion of SRC pairs. It is a measure
of the interaction of the pair with the `mean-field' potential created by the residual $A-2$ system.  Its magnitude,
as compared with the relative motion of the nucleons in the pairs, is key for establishing effective scale-separated
models of SRCs such as the GCF presented above and serves as an input for theoretical calculations.

The CM motion of SRC pairs is expected to be described by a gaussian distribution, defined by its width.
Therefore, experiments often report on their extraction of the C.M.~gaussian width, $\sigma_{CM}$.

\begin{figure}[htpb]
\centering
\includegraphics[width=0.6\textwidth]{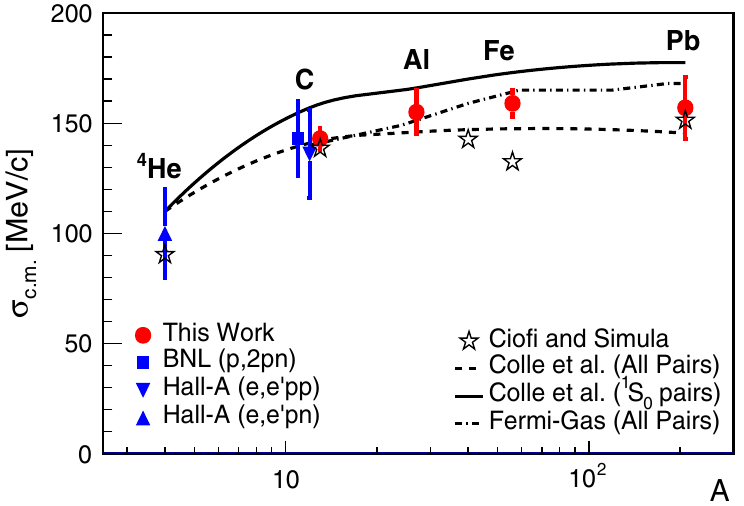}
\caption{\label{fig:erez_ppcm} 
Width of $pp$-SRC pairs C.M.~momentum distribution, extracted from $A(e,e'pp)$ data (red circles)~\cite{Cohen:2018gzh}, 
compared with previous extractions (blue points). The width is extracted assuming a 3D gaussian for the C.M.~distribution,
defined by its width, $\sigma_{CM}$. The lines and stars show mean-field theory predictions~\cite{CiofidegliAtti:1995qe,Colle:2013nna}.
}
\end{figure}

Fig.~\ref{fig:erez_ppcm} shows new results from the extraction of the $\sigma_{CM}$ for pp-SRC pairs from an analysis
of $A(e,e'pp)$ data, led by graduate student E.~Cohen and the spokespersons~\cite{Cohen:2018gzh}. 
The extracted C.M.~momentum distribution for the measured nuclei was observed to be consistent with a Gaussian
distribution in each direction, as expected. The extracted values of $\sigma_{CM}$ were observed to vary
between 140 and 160~MeV$/c$, and are consistent with a constant within experimental uncertainties. 

Comparisons with theory predictions show good agreement with either a simple Fermi-gas model prediction (where the $NN$ 
pairs are formed from two randomly chosen nucleons, each following a Fermi-Gas momentum distribution with $k_F = 250$~MeV$/c$)
or more realistic mean-field calculations~\cite{CiofidegliAtti:1995qe,Colle:2013nna}. Interestingly, the data seem to be higher
than the mean-field predictions that assume all $NN$ pairs can form SRC pairs, but lower than the most restrictive $^1$S$_0$
calculation (i.e., assuming only mean-field $pp$ pairs in a relative $^1$S$_0$ state can form $pp$-SRC pairs). This indicates
some selectivity in the SRC pair formation process and was suggested to provide insight to their 
quantum numbers~\cite{Cohen:2018gzh,Colle:2015ena,Colle:2013nna}.

\subsection{Final State Interactions in Hard QE Scattering}

\begin{figure}[htpb]
\centering
\includegraphics[height=6cm]{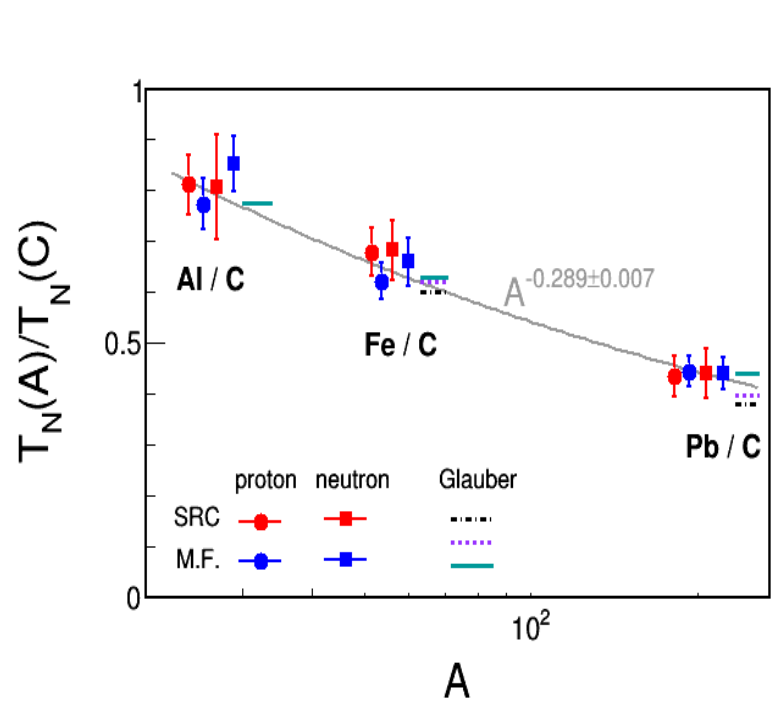}\hspace{8mm}
\includegraphics[height=6cm]{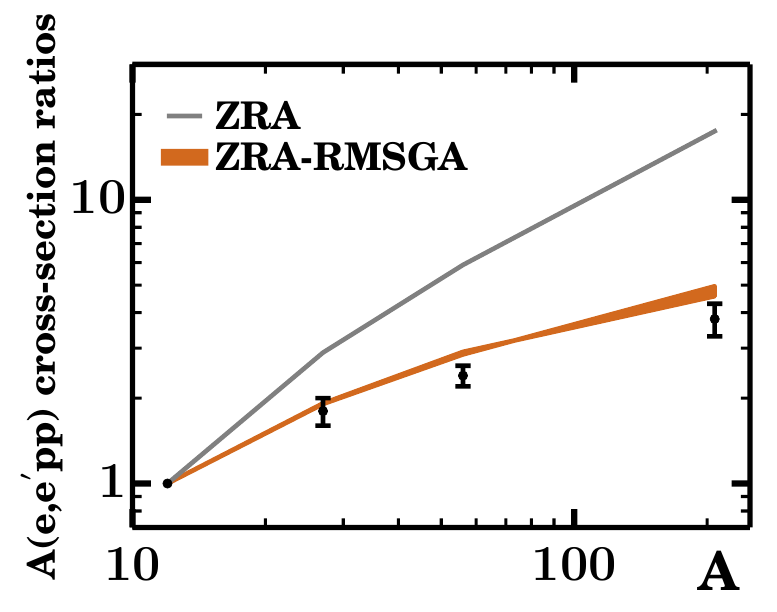}
\caption{\label{fig:reaction_mech}
Nucleon transparency ratios for nuclei relative to $^{12}$C, extracted from 
single-nucleon knockout measurements (left)~\cite{Duer:2018sjb}, and calculations of the
two-nucleon knockout reaction~\cite{Colle:2015ena} using Glauber theory (right).
}
\end{figure}

The results presented above in sections~\ref{ssec:gcf} and \ref{ssec:NN} require corrections for 
reaction effects such as final-state interactions (FSI) and singe-charge exchange (SCX). Therefore,
understanding the impact of such reaction mechanism effects on hard electron QE scattering cross-sections
is crucial for the interpretation of measurements in general, and specifically their relation to 
ground-state properties of nuclei. In high-$Q^2$ reactions, one may use the Generalized Eikonal 
approximation within a Glauber-framework to perform quantitative estimations of reaction effects
such as FSI and SCX. However, additional experimental verification of this approach in the kinematics
of our measurements are needed. Several measurements of the nuclear transparency of
proton knockout in $(e,e'p)$ and $(e,e'pp)$ reactions in SRC kinematics were compared them with
theoretical calculations using the Glauber approximation~\cite{Hen:2012yva,Colle:2015ena} (Fig.~\ref{fig:reaction_mech}, right).
The experimentally extracted transparency ratios showed good agreement with Glauber calculations.
Recently, this work was extended to measurements of neutron knockout $(e,e'n)$ reactions in both 
SRC and Mean-Field kinematics~\cite{Duer:2018sjb} (Fig. 7 top panel). The extracted transparency
for both proton and neutron knockout in mean-field and SRC kinematics were observed to agree with
each other and with Glauber calculations. The combined nuclear mass dependence of the data is
consistent with power-law scaling of $A^\alpha$ with $\alpha = -0.285 \pm 0.011$, which is 
consistent with nuclear surface dominance of the reactions.

\subsection{Nuclear Asymmetry Dependence in SRCs}

The predominance of $np$-SRC pairs leads to interesting phenomena in asymmetric nuclei. 
Without SRC pairs, neutrons in neutron-rich nuclei should have a higher Fermi momentum
and thus a higher average momentum and kinetic energy than the minority protons. However, 
since the high-momentum tail of the momentum distribution is dominated by $np$-pairs, 
there should be equal numbers of protons and neutrons above $k_F$. Therefore, the excess
neutrons in a neutron-rich nucleus should either increase the fraction of correlated 
protons or occupy low-momentum states. In either case, the fraction of high-momentum 
protons should be larger than that of neutrons~\cite{Hen:2014nza,Sargsian:2012sm,Ryckebusch:2018rct}.

\begin{figure}[htpb]
\centering
\includegraphics[height=5cm]{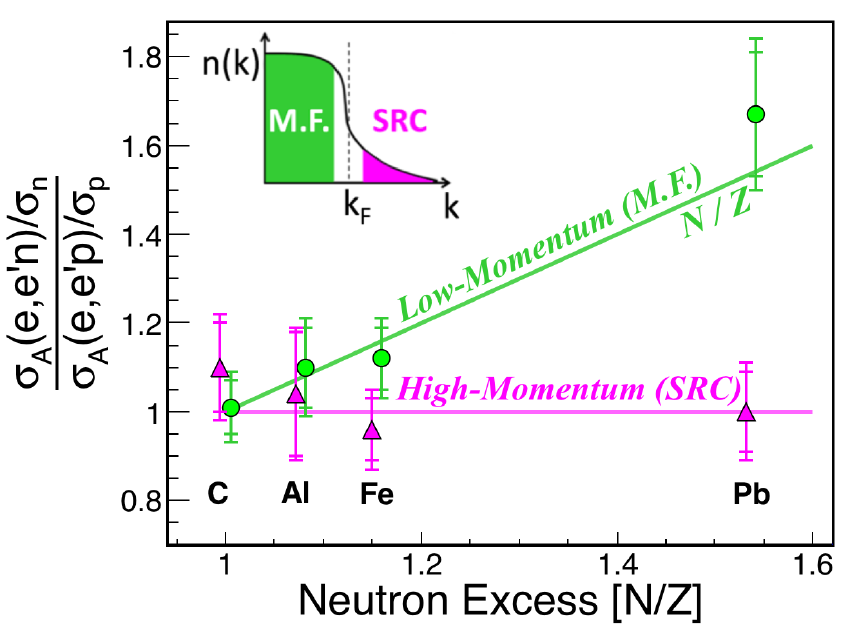}
\hspace{5mm}
\includegraphics[height=5cm]{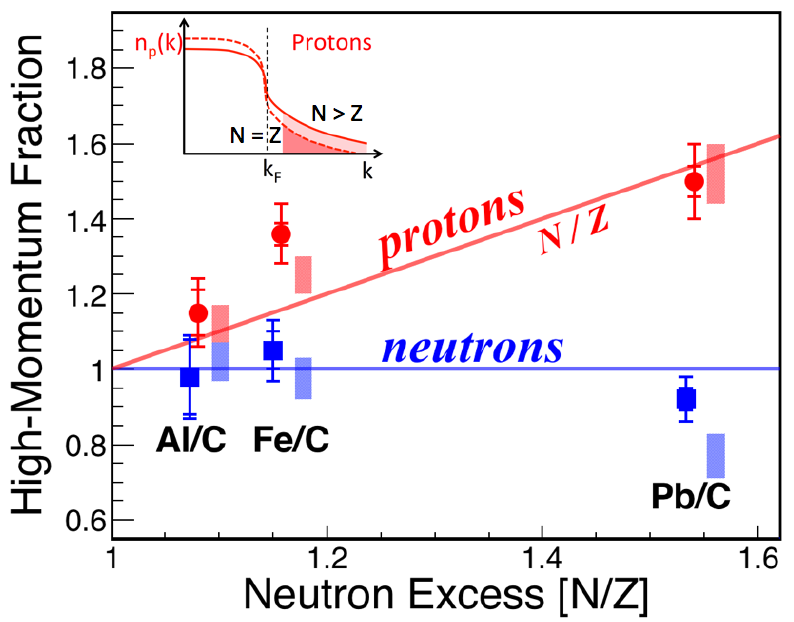}
\caption{\label{fig:meytal}
Nucleon knockout studies of heavy nuclei~\cite{Duer:2018sby}. Left: Extracted ratio of proton to neutron
knockout from above and below the nuclear Fermi momentum. Right: Extracted fraction of high-momentum $(k>k_F)$
protons and neutrons in nuclei relative to $^{12}$C, compared with SRC model predictions (shaded squares).
}
\end{figure}

In a paper recently published in Nature~\cite{Duer:2018sby}, we reported the first 
simultaneous measurement of hard QE electron scattering off protons and neutrons 
(i.e., $A(e,e'p)$ and $A(e,e'n)$ reactions) in $^{12}$C, $^{27}$Al, $^{56}$Fe, and 
$^{208}$Pb. The simultaneous measurement of both proton and neutron knockout allowed
a direct comparison of their properties with minimal assumptions. The measurement was
made in two different kinematical settings, one corresponding to electron scattering 
primarily off nucleons from an SRC pair ($p_\text{miss} > k_F$), the other from nucleons
in the nuclear mean field ($p_\text{miss} < k_F$). Using these event samples, the reduced
cross-section ratios: $[A(e,e'n)/\sigma_{en}] / [A(e,e'p)/\sigma_{ep}]$ (i.e., measured
cross-sections divided by the known elementary electron-proton, $\sigma_{ep}$, and 
electron-neutron, $\sigma_{en}$, cross-sections) were extracted for each kinematical setting.
The results shown in Fig. 8 (left) indicate that the $n/p$ mean-field reduced cross-section 
ratios grow approximately as $N/Z$ for all nuclei, as expected from simple nucleon counting.
However, the SRC ratios in all nuclei are consistent with unity, consistent with $np$-SRC
dominance of the high-momentum tail.

To quantify the pairing mechanism leading to constant $n/p$ ratios for SRC nucleons, 
we also extracted the relative fraction of high-missing-momentum to low-missing-momentum
events in neutron-rich nuclei relative to $^{12}$C, see Fig. 5 (right). This extraction
was done separately for protons and neutrons, and shows that the neutron SRC probabilities
are independent of the nuclear neutron excess (i.e. they saturate) while the corresponding
proton probabilities grow linearly with $N/Z$. This observation indicates that in
neutron-rich nuclei, the outer excess neutrons form SRC pairs with protons from the inner
`core' of the nucleus.

\subsection{Reaction Mechanisms Uncertainties in the Interpretation of SRCs}

\begin{figure}[htpb]
\centering
\includegraphics[height=3.5cm]{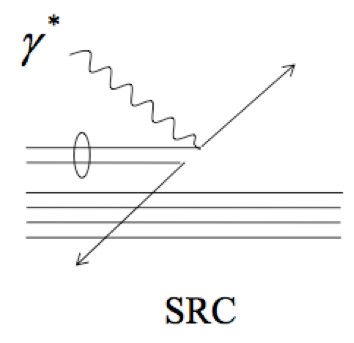}
\hspace{5mm}
\includegraphics[height=7cm]{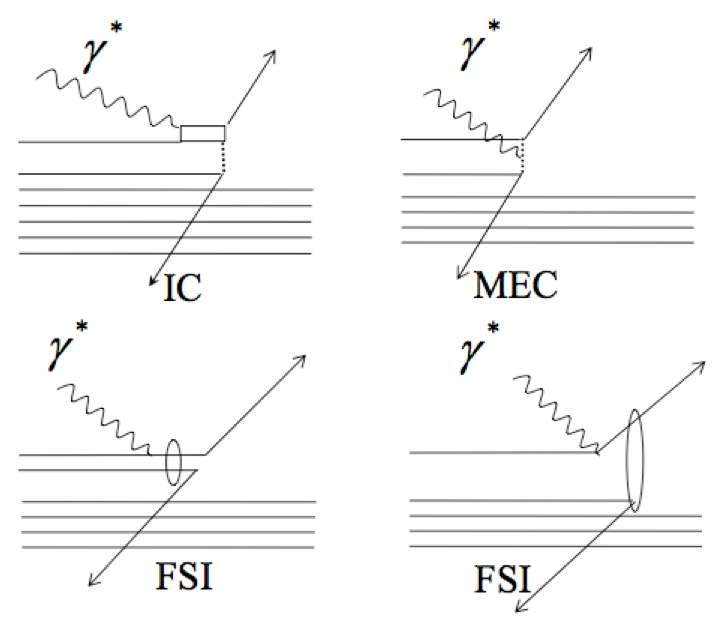}
\caption{\label{fig:ReacMech}
The reaction mechanisms for electron-induced two nucleon knockout.
The virtual photon can be absorbed on one nucleon of an SRC pair,
leading to the emission of both nucleons (SRC).  The virtual photon
can excite a nucleon to a $\Delta$, which deexcite by exchanging a
pion, resulting in the emission of two nucleons (IC).  The virtual
photon can be absorbed on a pion-in-flight (MEC).  The virtual photon
can be absorbed on one nucleon of an SRC pair which rescatters from
the other nucleon in the pair (FSI (left)).  The virtual photon
can be absorbed on an uncorrelated nucleon which rescatters from
another nucleon (FSI (right)).  }
\end{figure}

The results described above are almost all derived from electron scattering measurements, 
with only a single proton scattering C$(p,ppn)$ measurement~\cite{Tang:2002ww}.
Thus, the interpretation of these experimental results relies on an assumed electron
interaction mechanism at large momentum transfers. There are a number of different
electron-scattering reaction mechanisms that can lead to two-nucleon emission (see
Fig.~\ref{fig:ReacMech}).  While the experiments described above have been performed
at kinematics where many of these effects have been minimized, there are still
interpretational uncertainties due to these other possible reaction mechanisms.
These reaction mechanisms are not present or are very different for proton scattering.

\begin{figure}[htpb]
\centering
\includegraphics{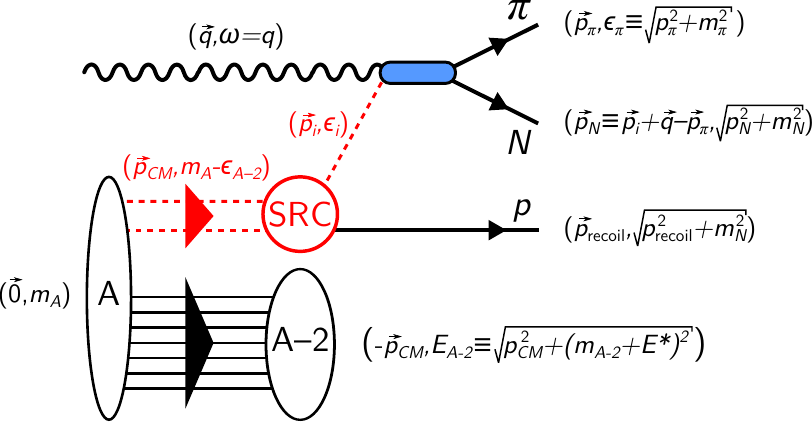}
\caption{\label{fig:reaction_g}
Diagrammatic representation and kinematics of the triple-coincidence $A(\gamma,\pi Np)$ reaction, one of the
main channels of interest for SRC breakup by a real photon beam. As in Fig.~\ref{fig:reaction_e}, dashed red
lines represent off-shell particles. Open ovals represent un-detected systems. Solid black lines represent 
detected particles. The momentum and energy of the particles are also indicated.}
\end{figure}

Photon scattering will also proceed through very different reaction mechanisms. 
Instead of quasielastic nucleon knockout, the primary photo-induced reaction studied
here will be $\gamma n\rightarrow p\pi^-$, with a second nucleon (the correlated
partner nucleon) emitted backward (see Fig.~\ref{fig:reaction_g}. For this reaction, the IC and MEC reaction
mechanisms will be absent or significantly different.  In addition, because the
correlated partner nucleon will be emitted backwards, the effects of Final State
Interactions (FSI) will also be quite different.  It is much more difficult to
produce backward nucleons that forward ones.

Thus photonuclear measurements of SRCs will provide a crucial reaction mechanism
check for SRC studies.

\section{Photonuclear probes of bound nucleon structure}

\label{sec:bns}

\subsection{The EMC Effect and SRCs}
The relative abundance of SRC pairs in nuclei can be extracted from measurements of inclusive 
$(e,e')$ cross-section ratios for different nuclei at high-$Q^2$, $x_B > 1$
kinematics~\cite{Hen:2016kwk,Atti:2015eda,Arrington:2011xs,Frankfurt:2008zv,Schmookler:2019nvf,Frankfurt:1993sp,Egiyan:2003vg,Egiyan:2005hs,Fomin:2011ng}.
For fixed $Q^2$, these cross-section ratios scale as a function of $x_B$ starting approximately
at $x_B \ge 1.5$ The height of the scaling plateau is often used to extract the relative number
of high-momentum nucleons (i.e. SRC pairs) in the measured nuclei. We refer to these as the `SRC
scaling coefficients'.

In a recent series of publications~\cite{Hen:2016kwk,Weinstein:2010rt,Hen:2012fm,Hen:2013oha},
we and others have shown that the extracted SRC scaling coefficients linearly correlate
with the strength of the EMC effect in nuclei from $^3$He to $^{197}$Au. The latter is the 
slope of the deviation from unity of the isoscalar DIS cross-section ratio for nuclei relative
to deuterium in the range $0.3 \le x_B \le 0.7$. The EMC effect is commonly interpreted as
evidence for modification of the partonic structure function of bound nucleons~\cite{Hen:2016kwk,CiofidegliAtti:2007ork,Kulagin:2010gd}.

The observation of a correlation between the strength of the EMC effect and the SRC scaling
coefficients in nuclei generated new interest in the EMC effect (see e.g. CERN Courier cover
paper from May 2013; `Deep in the nucleus: a puzzle revisited'~\cite{cern:courier}) and gave
new insight into its possible origin. Several models have been proposed by us and others that
attempt to explain the underlying dynamics that drive the EMC effect and its correlation with
SRC pair abundances; see a recent review in Ref.~\cite{Hen:2016kwk}.

\begin{figure}[htpb]
\centering
\includegraphics[width=\textwidth]{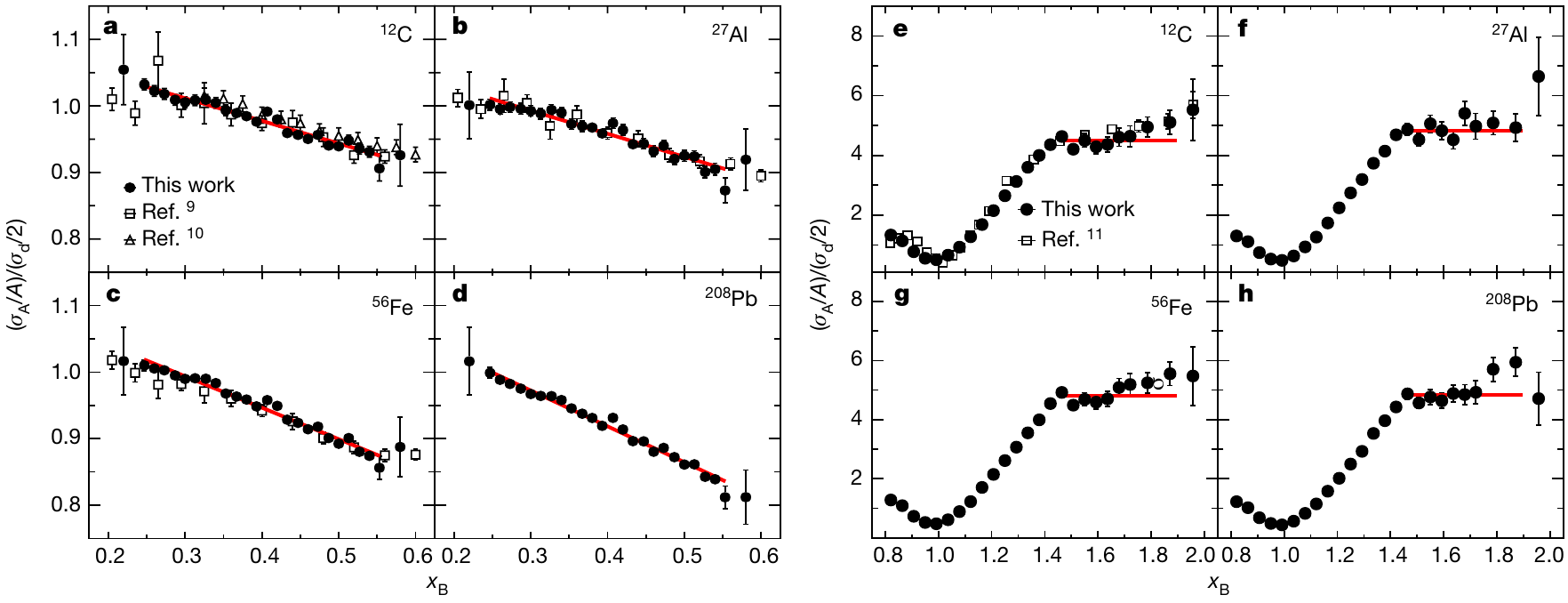}
\caption{\label{fig:barak_results}
High-precision measurements of the EMC effect (left) and SRC scaling (right) led by the spokespersons~\cite{Schmookler:2019nvf}.
}
\end{figure}

In a data-mining analysis recently published in Nature~\cite{Schmookler:2019nvf}, led by graduate 
student B.~Schmookler and the spokespersons, a high-precision measurement
of both the SRC scaling coefficients and the EMC effect was performed for $^{12}$C, $^{27}$Al, 
$^{56}$Fe and $^{208}$Pb (see Fig.~\ref{fig:barak_results}). The new data were used to examine
the finer aspects of the EMC-SRC correlation. Specifically, we examined whether the EMC data can
indeed be explained by assuming the nuclear structure function can be factorized into a collection
of un-modified mean-field nucleons and modified SRC pairs:
\begin{equation}
\label{eq:univ}
F_2^A = (Z - n_\text{SRC}^A) F_2^p + (N - n_\text{SRC}^A) F_2^n + n_\text{SRC}^A \left(F_2^{p*} + F_2^{n*}\right),
\end{equation}
where $n_\text{SRC}^A$ is the number of $np$-SRC pairs, $F_2^N(x_B)$ are the free nucleon (proton and neutron) structure
functions, and $F_2^{N*}(x_B)$ are the average modified nucleon structure functions in SRC pairs. $n_\text{SRC}^A$ 
is taken from experiment (i.e. from $(e,e')$ scaling ratios at $x_B > 1.5$), and the modified structure function of 
SRC nucleons, $F_2^{N*}(x_B)$, is expected to be universal (i.e., independent of the surrounding nuclear environment). 

\begin{figure}[htpb]
\centering
\includegraphics[width=0.8\textwidth]{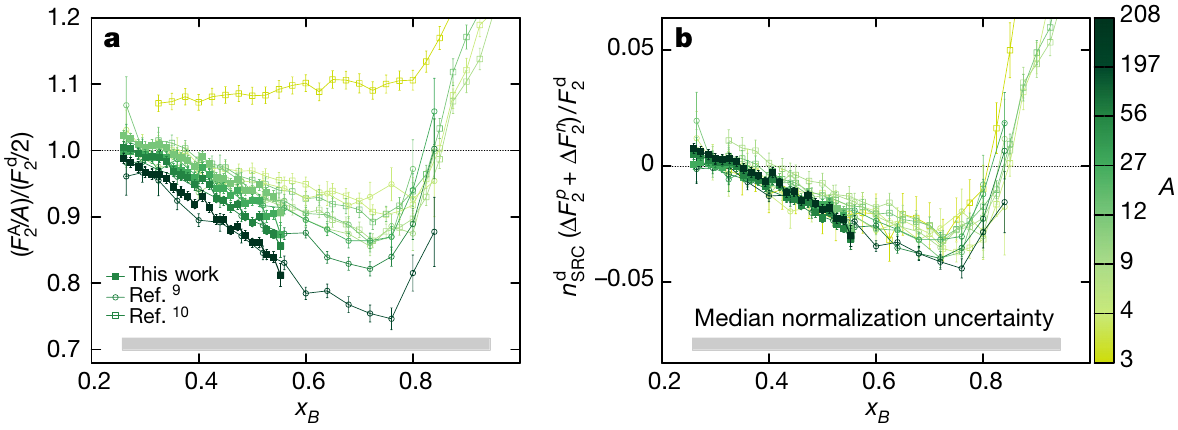}
\caption{\label{fig:barak_univ}
Left: measured structure function ratio for nuclei relative to deuterium 
(without model-dependent iso-scalar corrections).
Right: the extracted universal modification function of nucleons in SRC pairs~\cite{Schmookler:2019nvf}.
}
\end{figure}

Figure~\ref{fig:barak_univ} shows the measured structure function ratios of nuclei relative to
deuterium (left panel), and the extracted modification function of SRC pairs, using 
$\Delta F_2^N = F_2^{N*}-F_2^N$ (right panel). As can be seen, while the nuclear structure functions
vary significantly between different nuclei, the extracted SRC pair modification function is universal for all nuclei.

\subsection{Branching ratio modification}
To gain further insight to the modification of nucleons bound in SRC pairs, we propose to measure
the variation in the Branching Ratios (BRs) for hard photonuclear reactions off free (/quasifree)
vs. deeply bound nucleons in the deuteron, $^4$He and $^{12}$C. Changes in the measured BRs, which
may depend on the momentum transfer, scattering angle and nuclear transparency, will shed new light
on the mechanisms of quark-gluon nucleon structure modification in nuclei. The detailed description
of this novel observable follows.

The proton (or neutron) is a complex system that can be described in QCD at any given moment as a
superposition of different Fock states:
\begin{equation}
\label{eq:N_decomp}
\left| \text{proton} \right\rangle = \alpha_\text{PLC} \left|\text{PLC}\right\rangle 
+ \alpha_{3qg}\left| 3q+g\right\rangle + \alpha_{3q q\bar q} \left|3qq\bar q\right\rangle 
+ \alpha_{3q\pi}\left| 3q\pi \right\rangle + \ldots
\end{equation}
where the different brackets represent states of the proton (or neutron) with the corresponding $\alpha$
representing the amplitude of each state. By definition all weights must sum to 1. The minimal state of
the nucleon includes only the 3 valence quarks and is assumed to be small in size and with a reduced strong
interaction. Such a state is referred to as a Point-Like Configuration (PLC). The other states include more
complex configuration involving additional gluons, quark-antiquark pairs, pions etc. {\it These states are
all components of the wave function of the nucleon.}

The modified structure of a proton (or neutron) bound in a nucleus can then be represented by
a different, decomposition into the same Fock states:
\begin{equation}
\label{eq:N_decomp_bound}
\left| \text{proton}^* \right\rangle = \alpha_\text{PLC}^* \left|\text{PLC}\right\rangle 
+ \alpha_{3qg}^* \left| 3q+g\right\rangle + \alpha_{3q q\bar q}^* \left|3qq\bar q\right\rangle 
+ \alpha_{3q\pi}^* \left| 3q\pi \right\rangle + \ldots
\end{equation}
where the difference between a free and bound proton is depicted by the difference between the $\alpha$ and $\alpha^*$
coefficients in Eqs.~\ref{eq:N_decomp} and \ref{eq:N_decomp_bound}. An example of such an effect can be found in the 
`Point Line Configurations Suppression' model of Frankfurt and Strikman~\cite{Frankfurt:1985cv} or the `Blob-Like 
Configurations Enhancement' model of Frank, Jennings, and Miller~\cite{Frank:1995pv} that propose a possible explanation
to the EMC effect in which the PLC part of the bound nucleon is different than in a free one.

We stress that the Fock space description of bound nucleons is somewhat more complex as nucleons bound in nuclei
span various states: e.g. mean-field vs. SRC nucleons, high vs. low local density etc., allowing the $\alpha^*$
coefficients to possibly depend on the detailed nuclear state of the bound nucleon. For example, in the PLC 
suppression model~\cite{Frankfurt:1985cv}, $|\alpha_\text{PLC}^*/ \alpha_\text{PLC}|^2 -1$ is proportional to
the nucleon off-shellness (approximately to the square of the nucleon momentum) with much smaller modification
for configurations close to the average ones. Hence one expects maximal bound nucleon modification in the 
processes dominated by scattering from small size configurations.

We expect that different Fock states will absorb high-energy photons differently and lead to different branching
ratios (BR) for various final states (e.g., $\gamma p \rightarrow \pi^- \Delta^{++}; \rho^0 p$). We propose to use the unique
capability of GlueX to measure simultaneously the BRs of many decay channels of an excited nucleon following the
absorption of a real photon at high momentum transfer (large t). By measuring these BR for nucleons in a range of
nuclei from deuterium through lead we will be able to see differences in the Fock state decomposition, and hence 
the structure, of bound and free nucleons.

\begin{table}
\centering
\caption{\label{table:reactions} List of possible exclusive photonuclear reactions off protons
and neutrons that are within the detection capabilities of the GlueX spectrometer. Note that
neutron reactions are only possible using nuclear targets (deuteron and heavier).}
\vspace{2pt}
\begin{tabular}{| l | l | }
\hline
\textbf{Proton Reactions} & \textbf{Neutron Reactions} \T \B \\
\hline
$\gamma + p \rightarrow \pi^0 + p $ & $\gamma + n \rightarrow \pi^- + p $ \T \B \\
\hline
$\gamma + p \rightarrow \pi^- + \Delta^{++}$ & $\gamma + n \rightarrow \pi^- + \Delta^+$ \T \B \\
\hline
$\gamma + p \rightarrow \rho^0 + p $ & $\gamma + n \rightarrow \rho^- + p$ \T \B \\
\hline
$\gamma + p \rightarrow K^+ + \Lambda^0$ & $\gamma + n \rightarrow K^0 + \Lambda^0$ \T \B \\
\hline
$\gamma + p \rightarrow K^+ + \Sigma^0$ & $\gamma + n \rightarrow K^0 + \Sigma^0$ \T \B \\
\hline
$\gamma + p \rightarrow \omega + p $ & --- \T \B \\
\hline
$\gamma + p \rightarrow \phi + p $ & --- \T \B \\
\hline
\end{tabular}
\end{table}

For a free proton GlueX will measure the branching ratio (BR) for many reactions, including 
$\gamma p \rightarrow \pi^- \Delta^{++}$, $\rho^0 p$, $K^+ \Lambda$, $K^+ \Sigma^0$, and others.
By measuring these reactions, and the neutron equivalents (listed in Table~\ref{table:reactions}), 
on deuteron and nuclear targets, we can extract the BRs for scattering off free (/quasi-free) vs. deeply bound nucleons.
As each reaction is sensitive to a different combination of Fock states, modifying their contribution to
the bound proton will modify their BRs.

Current theoretical models do not allow us either to predict the exact change in BRs as a function
of the bound nucleon structure, or to translate the observable BRs to the modified $\alpha^*$ coefficients.
However, this is a novel observable that allows us to observe or exclude deviations, and to study their
dependence on the nucleon momentum and `hardness' of the reaction. Any such observation will therefore
serve as clear and direct evidence for changes in bound nucleon structure.

On average, differences between a bound and a free nucleon are expected to be small. However, we propose
to select specific kinematics, focusing on deeply bound nucleons, that could enhance the effect: e.g.,
selecting hard process with large $s$, $t$, and $u$ is expected to emphasize the contribution of the PLC
component. Alternatively, detecting the decay products along with a high-momentum recoil nucleon (which
favors scattering from a nucleon from an SRC pair) should also significantly amplify the medium effect
to the level observed for the EMC effect at $x_B \sim 0.5$--$0.6$ that is modification on the scale of 20\%.
The high-momentum recoil nucleons can best be studied with a deuteron target.

In the case of nuclear targets, the measured BRs are not at the photon absorption point but rather
following hadron attenuation in the nucleus, which may be different for each channel. To extract
the BR at the hard vertex of the quasi-free scattering we need to correct for hadron attenuation
in the nucleus. We propose to do that by measuring the process on deuterium, helium and carbon.
This will yield, as a byproduct, an interesting study of photon and Color Transparency (CT), 
as well as measure pattern of interaction with media of different mesons. See separate discussion on CT below.

\section{CT at High Energies}
\label{sec:ct}
In addition to SRCs and bound nucleon structure, the data to be collected in this 
experiment will be used to study hadron color transparency (CT).
At high energies, the phenomena of CT arises from the fact that exclusive processes
on a nucleus at high momentum-transfer preferentially select the color singlet, small 
transverse size configuration, which then moves with high momentum through the nucleus. 
The interactions between the small transverse size configuration and other nucleons are
strongly suppressed because the gluon emission amplitudes arising from different quarks 
cancel. This suppression of the interactions is one of the essential ingredients needed 
to account for Bjorken scaling in deep-inelastic scattering at small $x_B$ \cite{Frankfurt:1988nt}. 

CT at high energies was directly observed in the diffractive dissociation of 500~GeV$/c$ 
pions into dijets when coherently scattering from carbon and platinum targets. The 
per-nucleon cross-section for dijet production is parametrized as $s = s_0 A^a$, and 
the experiment found $a = 1.61 \pm 0.08$ \cite{Aitala:2000hb}, consistent with CT 
predictions of $a = 1.54$ \cite{Frankfurt:1993it}. These results confirm the predicted 
strong increase of the cross-section with $A$, and the dependence of the cross-section
on the transverse momentum of each jet with respect to the beam axis ($k_t$) indicates
the preferential selection of the small transverse size configurations in the projectile.
Such experiments have unambiguously established the presence of small-size $q\bar{q}$ Fock 
components in light mesons and show that at transverse separations, $d \sim 0.3$~fm, 
pQCD reasonably describes small $q\bar{q}$-dipole-nucleon interactions. Thus, color 
transparency is well established at high energies and low $x_B$. However, these high-energy 
experiments do not provide any information about the appropriate energy regime for the onset of CT.

At intermediate energies, in addition to the preferential selection of the small-size configuration, 
the expansion of the interacting small-size configuration is also very important. At these energies, 
the expansion distance scales are not large enough for the small-size configuration to escape 
without interaction which, suppresses the color transparency effect \cite{Farrar:1988me,Jennings:1989hc,Jennings:1990ma,Jennings:1992hs}.
The interplay between the selection of the small transverse size and its subsequent expansion 
determine the scale of the momentum and energy transfers required for the onset of CT. As 
mentioned, a major difference between photo-induced and electron-induced reactions is that,
in the former, much greater energy is transferred relative to momentum, which can help
disentangle the roles of freezing and squeezing.

The first attempt to measure the onset of CT at intermediate energies used the large-angle 
$A(p,2p)$ reaction at the Brookhaven National Lab (BNL) \cite{Carroll:1988rp,Mardor:1998zf,Leksanov:2001ui,Aclander:2004zm}.
In these experiments, large-angle $pp$ and quasielastic $(p,2p)$ scattering were 
simultaneously measured in hydrogen and several nuclear targets, at incident proton 
momenta of 6--12~GeV$/c$. The nuclear transparency was extracted from the ratio of 
the quasielastic cross-section from a nuclear target to the free $pp$ elastic cross-section. 
The transparency was found to increase as predicted by CT, but only between 6--9.5~GeV$/c$;
the transparency was found to decrease between 9.5 and 14.4~GeV$/c$. This decrease cannot be explained by models 
incorporating CT effects alone. Though not fully understood to date, this behavior is commonly 
attributed to a lack of understanding of the fundamental two-body reactions, which limits 
one’s ability to relate the $s$, $t$ scales for the onset of squeezing in different 
reactions. This situation raises doubts about our ability to study CT effects using 
such proton-induced QE scattering reactions.

In contrast to hadronic probes, weaker electromagnetic probes sample the complete 
nuclear volume. The fundamental electron-proton scattering cross-section is smoothly
varying and is accurately known over a wide kinematic range. Detailed knowledge of 
the nucleon energy and momentum distributions inside a variety of nuclei have been 
extracted from extensive measurements in low-energy electron scattering experiments. 
Therefore, the $(e,e'p)$ reaction is simpler to understand than the 
$(p,2p)$ reaction, an advantage immediately recognized following the BNL $(p,2p)$ 
experiments. A number of $A(e,e'p)$ experiments have been carried out over the years, 
first at SLAC \cite{Makins:1994mm,ONeill:1994znv} and later at JLab \cite{Abbott:1997bc,Garrow:2001di} for a range of 
light and heavy nuclei. In high $Q^2$ quasielastic 
$(e,e'p)$ scattering from nuclei, the electron scatters preferably from a single 
proton, which need not be stationary due to Fermi motion \cite{Frullani}. In the plane wave 
impulse approximation (PWIA), the proton is ejected without final state interactions
with the residual $A-1$ nucleons. The measured $A(e,e'p)$ cross-section would be 
reduced compared to the PWIA prediction in the presence of final state interactions, 
where the proton can scatter both elastically and inelastically from the surrounding 
nucleons as it exits the nucleus. The deviations from the simple PWIA expectation is 
used as a measure of the nuclear transparency. In the limit of complete color 
transparency, the final state interactions would vanish and the nuclear transparency
would approach unity. In the conventional nuclear physics picture, one expects the 
nuclear transparency to show the same energy dependence as the energy dependence of 
the $NN$ cross-section. Other effects such as short-range correlations and the density
dependence of the $NN$ cross-section will affect the absolute magnitude of the nuclear
transparency but have little influence on the energy- (or $Q^2$-) dependence of the 
transparency. Thus, the onset of CT is expected to be manifested as a rise in the
nuclear transparency as a function of increasing $Q^2$.

The existing world data rule out any onset of CT effects larger than 7\% over the
$Q^2$ range of 2.0--8.1~(GeV$/c$)$^2$ with a confidence level of at least 90\%. 
As mentioned earlier, the onset of CT depends both on momentum and energy transfers,
which affect the squeezing and freezing respectively. Since $A(e,e'p)$ scattering
measurements are carried out at $x_B=1$ kinematics, they are characterized by lower
energy transfers as compared to the momentum transfer (e.g. 4.2~GeV for $Q^2=8$~GeV$^2$). 
Existing data seem to suggest that a $Q^2$ of 8~(GeV$/c$)$^2$ with 4.2~GeV energy 
transfer is not enough to overcome the expansion of the small transverse size
objects selected in the hard $ep$ scattering process (i.e. freezing requirements 
are not met). A recent Hall C 12~GeV experiment, currently under analysis, 
will extend these studies to $Q^2 \sim 16$~GeV$^2$ \cite{E12:06:107}.
Although, no unambiguous evidence for CT has been observed so 
far for nucleons from either $A(e,e'p)$ or $A(p,2p)$ reactions, it is 
expected to be more probable to reach the CT regime at low energy for the
interaction/production of mesons than for baryons, since only two quarks must come
close together and a since a quark-antiquark pair is more likely to form a small size object 
\cite{Blaettel:1993rd}. Indeed, pion production measurements at JLab reported evidence
for the onset of CT \cite{Clasie:2007aa} in the process $e + A \rightarrow e + p + A^{*}$. 
The pion-nuclear transparency was calculated as the ratio of pion electroproduction 
cross-section from the nuclear target to that from the deuteron. As proposed here, 
the use of the deuteron instead of the proton helped reduce the uncertainty due to the unknown
elementary pion electroproduction cross-section off a free neutron and to uncertainties in the Fermi smearing
corrections. The measured pion nuclear transparency shows a steady rise with increasing pion 
momentum for the $A>2$ targets, and this rise in nuclear transparency versus $p_\pi$ is 
consistent with the rise in transparency predicted by various CT calculations 
\cite{Larson:2006ge,Cosyn:2006vm,Kaskulov:2008ej}. Although, all the calculations use an
effective interaction based on the quantum diffusion model \cite{Farrar:1988me} to 
incorporate the CT effect, the underlying conventional nuclear physics is calculated
very differently. The results of the pion electroproduction experiment demonstrate that
both the energy and $A$ dependence of the nuclear transparency show a significant deviation
from the expectations of conventional nuclear physics and are consistent with calculations
that include CT. The results indicate that the energy scale for the onset of CT in mesons
is $\sim 1$~GeV.

\begin{figure}[htpb]
\centering
\begin{minipage}{.49\textwidth}
\centering
\includegraphics[width=0.99\textwidth]{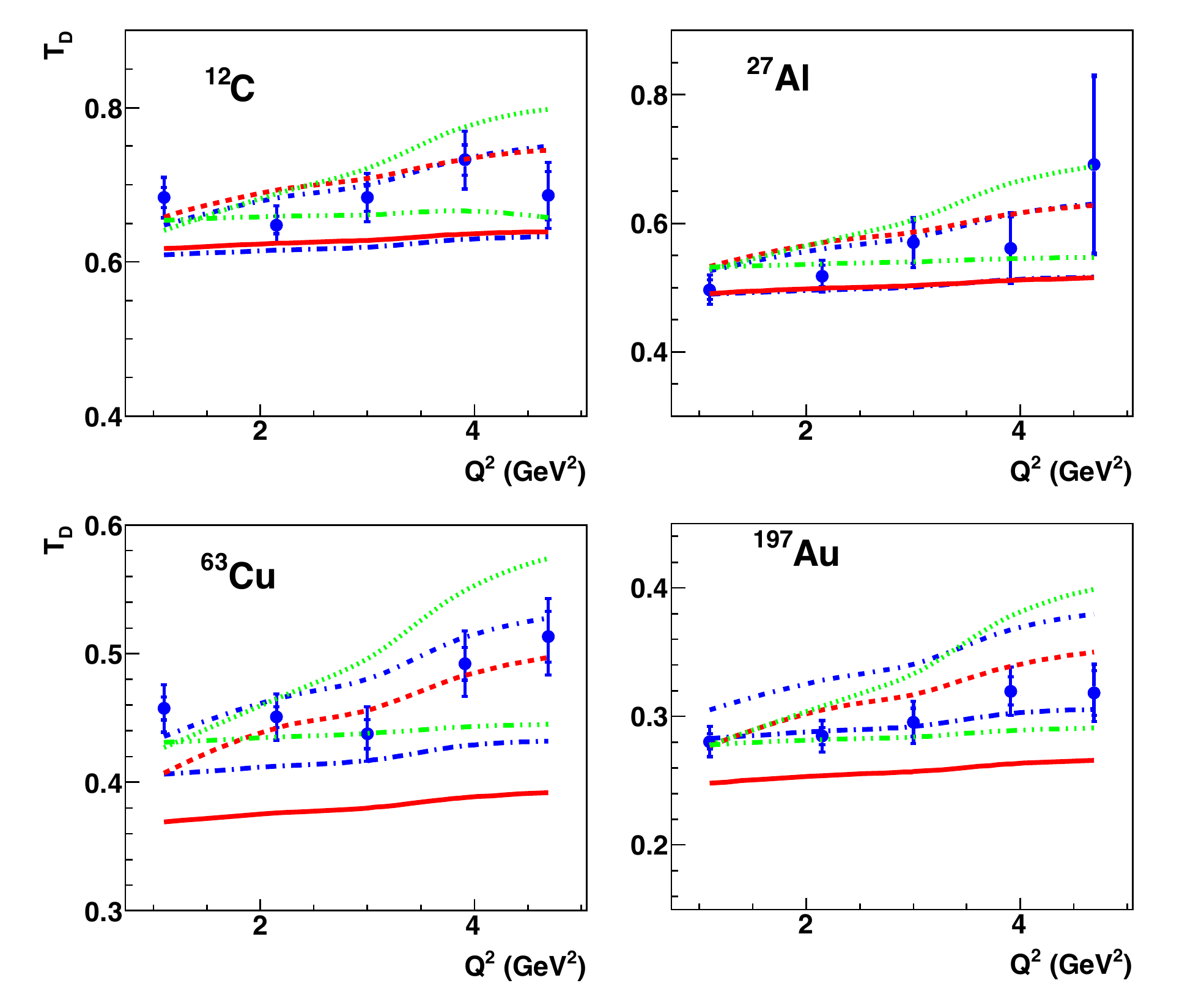}
\end{minipage}
\begin{minipage}{.49\textwidth}
\centering
\includegraphics[width=0.88\textwidth]{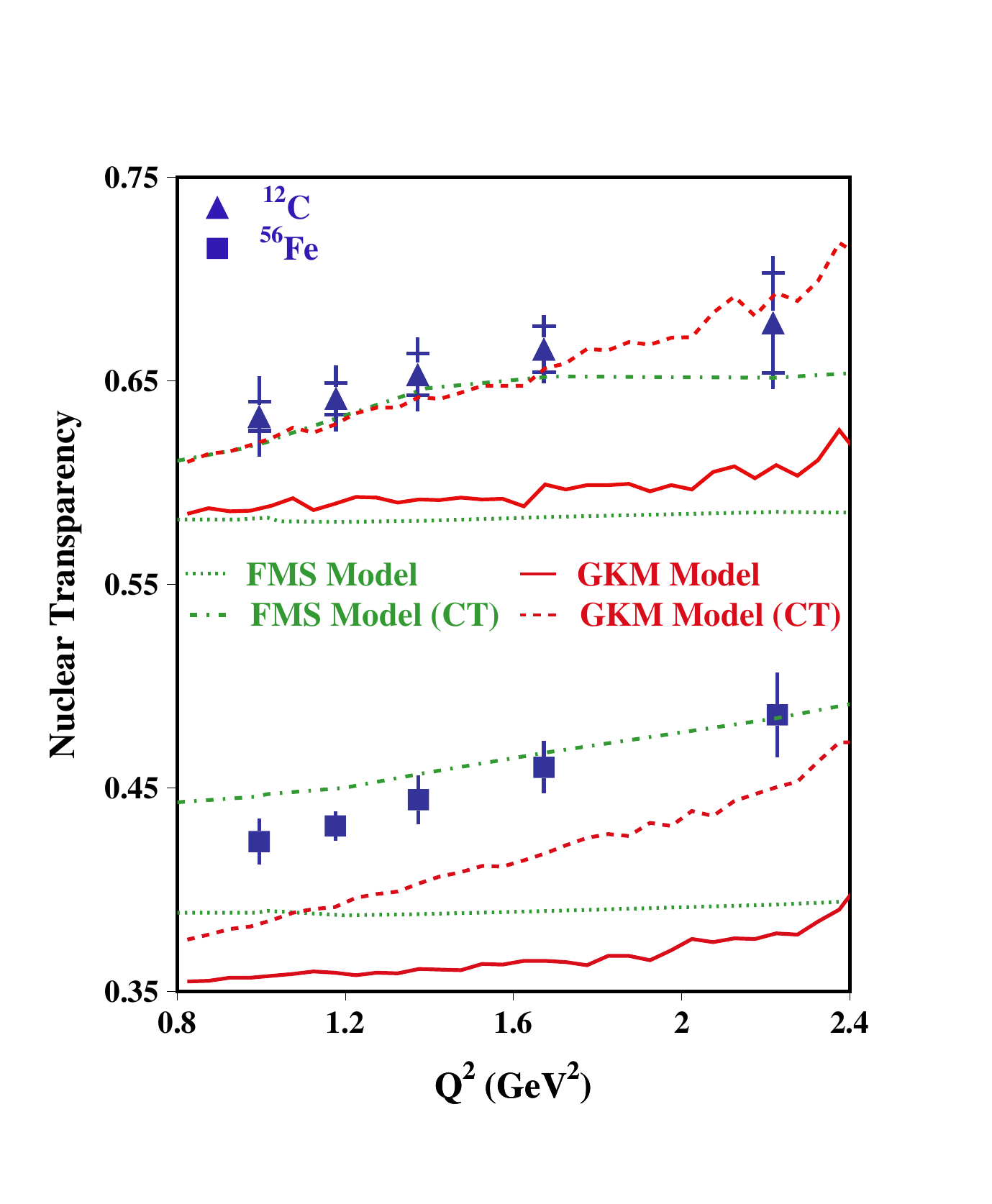}
\end{minipage}
\caption{\label{fig:clas_q2}
The two JLab experiments which show conclusive evidence for the onset of CT in meson electroproduction 
at intermediate energies. \textbf{(left panel)} Nuclear transparency vs $Q^2$ for $^{12}$C, $^{27}$Al, 
$^{63}$Cu and $^{197}$Au in the $(e,e^\prime \pi^+)$ reaction. The inner error bars are the statistical uncertainties and the outer error bars 
are the statistical and point-to-point systematic uncertainties added in quadrature. The solid circles 
(blue) are the high-$\epsilon$ (virtual photon polarization) points, while the solid squares (red) are 
the low-$\epsilon$ points. The dashed and solid lines (red) are Glauber calculations from \cite{Larson:2006ge}, 
with and without CT, respectively. Similarly, the dot-short dash and dot-long dash lines (blue) are Glauber 
calculations with and without CT from \cite{Cosyn:2006vm}. The dotted and dot-dot-dashed lines (green) are 
microscopic+ BUU transport calculations from \cite{Kaskulov:2008ej}, with and without CT, respectively. 
\textbf{(right panel)} Nuclear transparency as a function of $Q^2$ in the $(e,e^\prime \rho^0)$ reaction. The curves are predictions of the FMS 
\cite{Frankfurt:2008pz} (red) and GKM \cite{Gallmeister:2010wn} (green) models with (dashed-dotted and 
dashed curves, respectively) and without (dotted and solid curves, respectively) CT. Both models include 
the pion absorption effect when the $\rho^0$ meson decays inside the nucleus. The inner error bars are the 
statistical uncertainties and the outer ones are the statistical and the point-to-point systematic 
uncertainties added in quadrature.
}
\end{figure}

Electroproduction of vector mesons from nuclei is another excellent tool to investigate the formation
and propagation of quark-antiquark ($q\bar{q}$) pairs under well-controlled kinematical conditions. Soon
after the observation of the onset of CT in pion electroproduction, results from a study of $\rho$-meson 
production from nuclei at JLab also indicated an early onset of CT in mesons \cite{ElFassi:2012nr}. 
Previous $\rho^0$ production experiments had shown that nuclear transparency also depends on the coherence
length, $l_c$, which is the length scale over which the $q\bar{q}$ states of mass $M_{q\bar{q}}$ can propagate.
Therefore, to unambiguously identify the CT signal, one should keep $l_c$ fixed while measuring the $Q^2$ 
dependence of the nuclear transparency. The CLAS collaboration at JLab measured the nuclear transparency for 
incoherent exclusive $\rho^0$ electroproduction off carbon and iron relative to deuterium \cite{ElFassi:2012nr}
using a 5~GeV electron beam. An increase of the transparency with $Q^2$ for both C and Fe, was observed 
indicating the onset of CT phenomenon. The rise in transparency was found to be consistent with predictions 
of CT by models \cite{Frankfurt:2008pz,Gallmeister:2010wn} which had successfully described the increase in 
transparency for pion electroproduction. Therefore, the $\pi$ and $\rho$ electroproduction data also demonstrate
an onset of CT in the few GeV energy range as shown in Figure \ref{fig:clas_q2}. Both of these experiments will be extended to 
higher energies in future 12~GeV experiments \cite{E12:06:106,E12:06:107}.

\begin{figure}[htpb]
\centering
\includegraphics[width=0.49\textwidth]{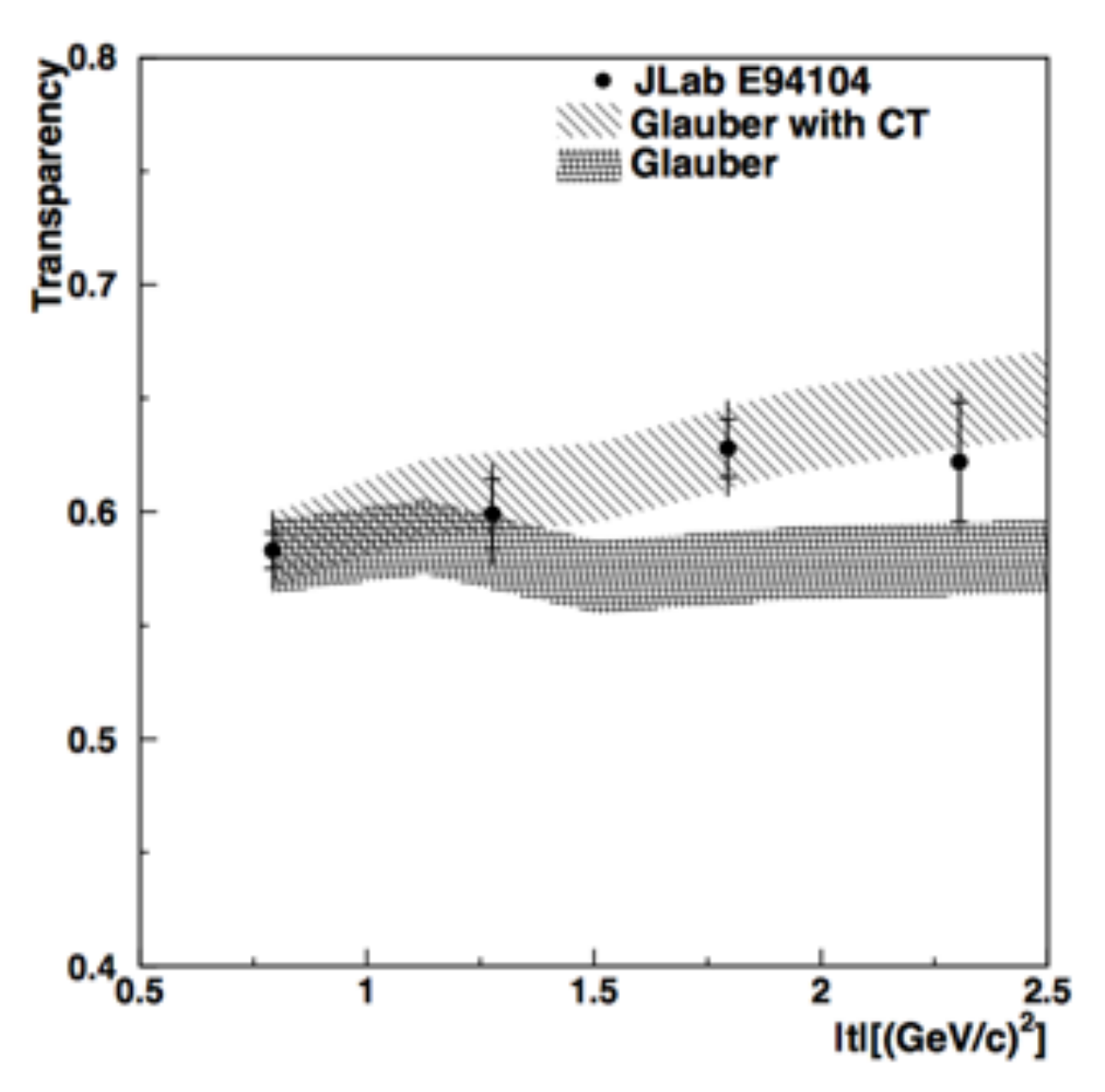}
\includegraphics[width=0.49\textwidth]{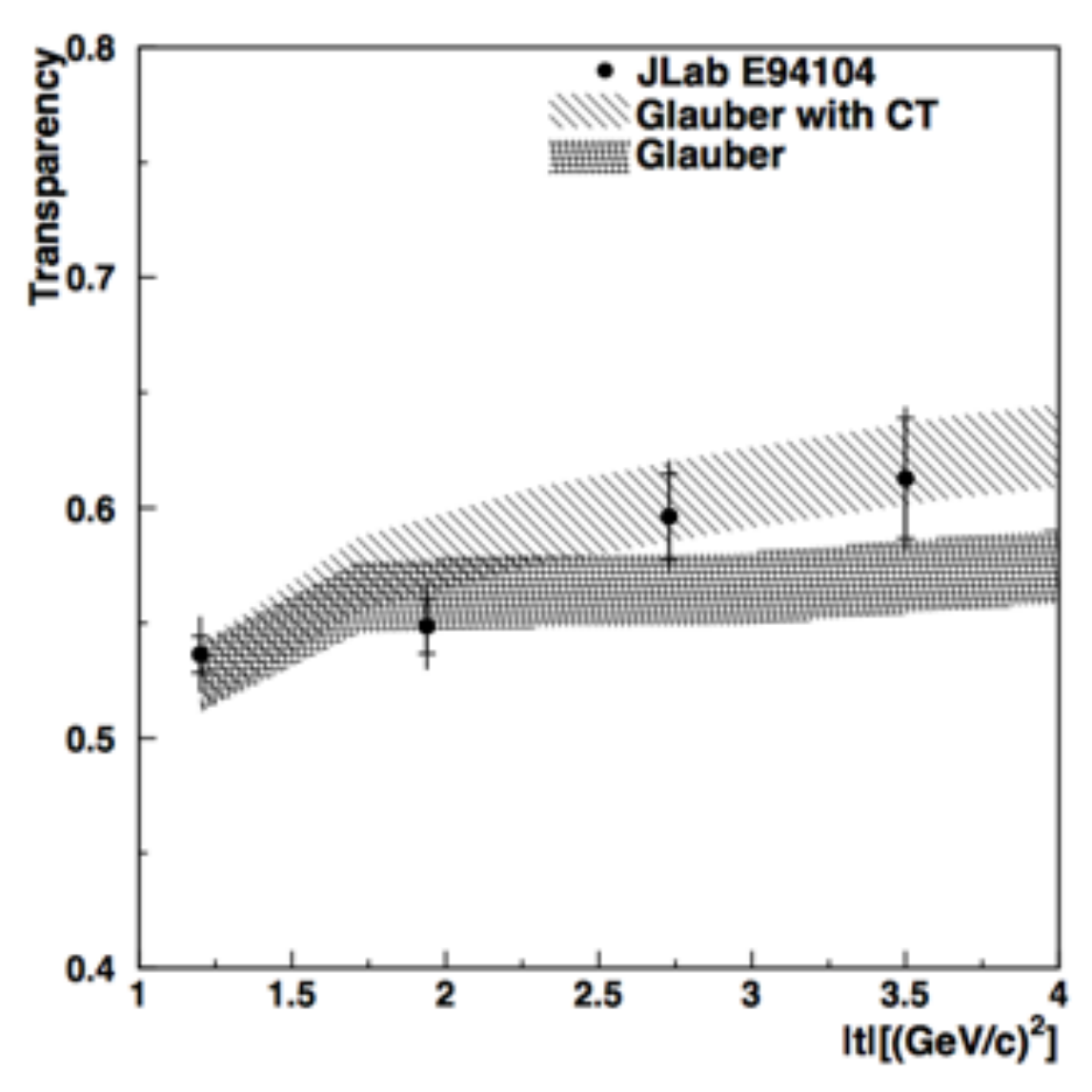}
\caption{\label{fig:ct}
The measured transparency for the reaction $^4\mbox{He}(\gamma,\pi^-p)$ for two c.m. scattering angles ($70^\circ$-right, $90^\circ$-left). 
The measurements are compared to Glauber calculations with and without CT. See Ref.\ \cite{Frankfurt:2008pz}
for details.
}
\end{figure}

In the case of large momentum transfer exclusive photoinduced reactions, while the predicted effects are larger 
(Fig.~\ref{fig:fig:rates_trans}) they were not studied in much detail. JLab experiment E94-104 
searched for CT using the reaction $\gamma + n \rightarrow \pi^- + p$~\cite{Dutta:2003mk}. The experiment used 
an untagged mixed electron and photon bremsstrahlung beam incident on a $^4$He target and the Hall-A high-resolution 
spectrometers to measure $\pi^-$ and $p$ produced in the reaction. The momentum transfer was reconstructed assuming
scattering off a mean-field neutron in $^4$He leaving the residual system in the ground state of $^3$He. Nuclear 
transparency was measured as a ratio of the pion photoproduction cross-section from $^4$He to that of $^2$H. Figure \ref{fig:ct}
shows the extracted transparency as a function of the momentum transfer, $|t|$, for center-of-mass
scattering angles of $70^\circ$ and $90^\circ$. The results were compared to Glauber calculations with and without 
CT effects. As can be seen, the measurement did not have the required statistical and systematical accuracy to 
discriminate between the two calculations over the measured $|t|$ range. We propose, using the advantages of the GlueX
spectrometer and the upgraded CEBAF 12~GeV electron beam, to add many more reaction channels, extend the measured 
$|t|$ range up to 10--12~GeV$^2$, and add heavier nuclei. While we will keep comparable uncertainties, for heavier
nuclei and larger momentum transfers the expected effects are considerably larger, which significantly increases the
discovery potential.

It should be pointed out that the rate of expansion/contraction of configurations involved in the interaction
with nucleons is the same for the different reactions. Hence, in light of the successful description of CT for mesons,
reliable estimate of space-time evolution effects were performed for other reactions with the conclusion that in 
the proposed kinematics for GlueX, CT is not washed out by the expansion/contraction effects due to the high photon
energies used in the experiment.

\section{The Proposed Measurement}

\subsection{Kinematics}

\label{ssec:kinematics}

The kinematical distributions and expected event rates were simulated for the pion-proton photo-production reaction 
off a neutron bound in a nucleus, $A(\gamma,\pi^- p)$, using a dedicated Monte-Carlo event generator. In this section, 
we present the simulation method and show the resulting kinematical distributions. 

\begin{figure}[htpb]
\centering
\includegraphics[width=0.8\textwidth]{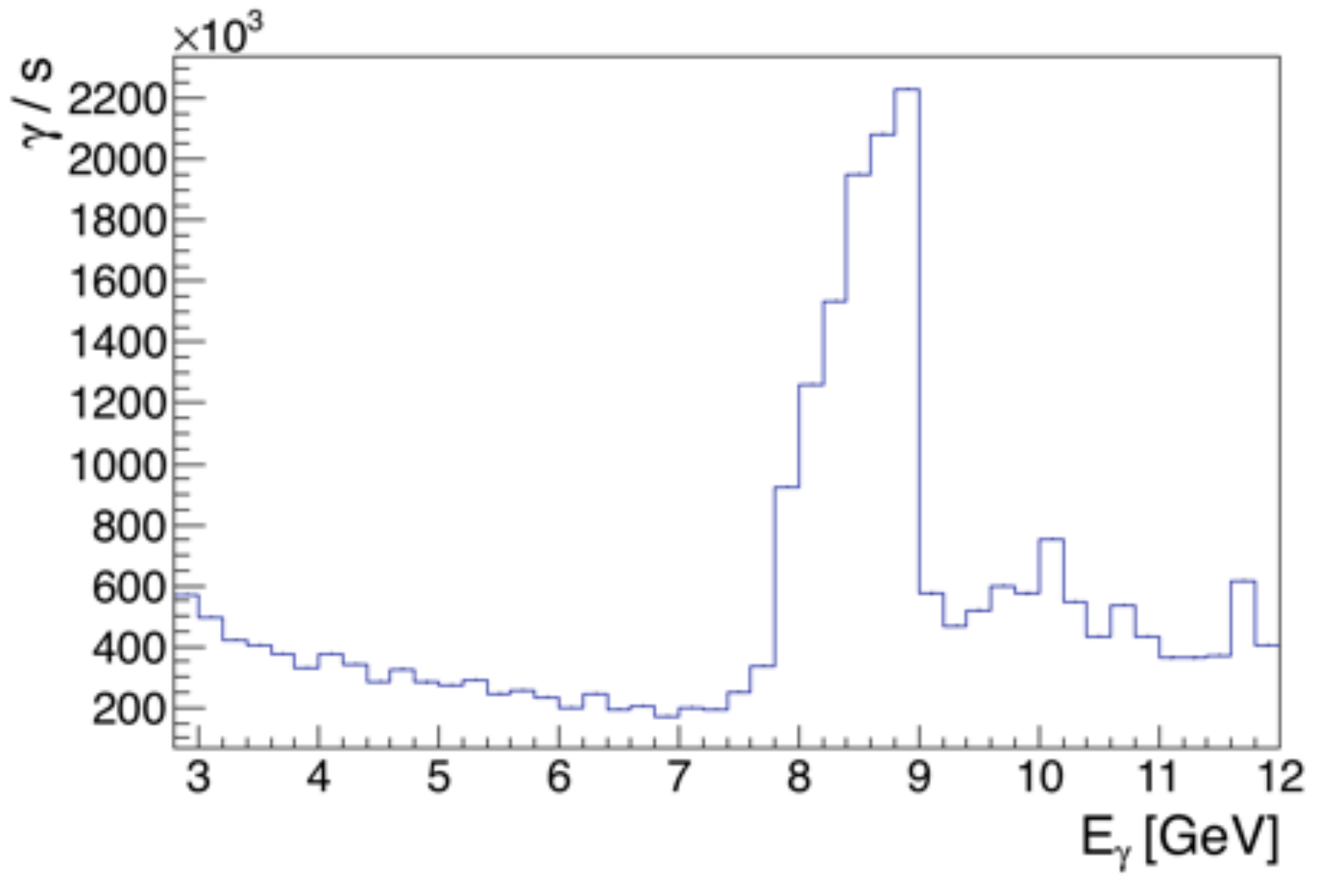}
\caption{\label{fig:gluex_e}
The energy distribution for the incoming photon beam hitting the GlueX target assuming a 5~mm diameter collimator. 
The distribution is normalized to a flux of $2.5\times 10^7$ photons/s in the beam energy range 7.5~GeV~$< E_\text{beam} <$~11.7~GeV.
}
\end{figure}

The simulation uses an incoming photon with energy sampled from the tagged photon spectra obtained from the standard 
GlueX simulation software (Fig.\ \ref{fig:gluex_e}). The momentum distribution of the nucleons in the nucleus has two 
components: a mean-field region that spans low momentum (up to $k_F$) and account for 80\% of the nucleons and an SRC
region that spans high momentum (from $k_F$ and up) and account for 20\% of the nucleons. The SRC-pairs are modeled
using a three-dimensional Gaussian center of mass momentum distribution with width (sigma) of 140~MeV$/c$ 
\cite{Shneor:2007tu,Tang:2002ww,Korover:2014dma,CiofidegliAtti:1995qe,Colle:2013nna} In the case of the deuteron, the AV18 
momentum distribution was used.

\begin{figure}[htpb]
\centering
\includegraphics[width=0.54\textwidth]{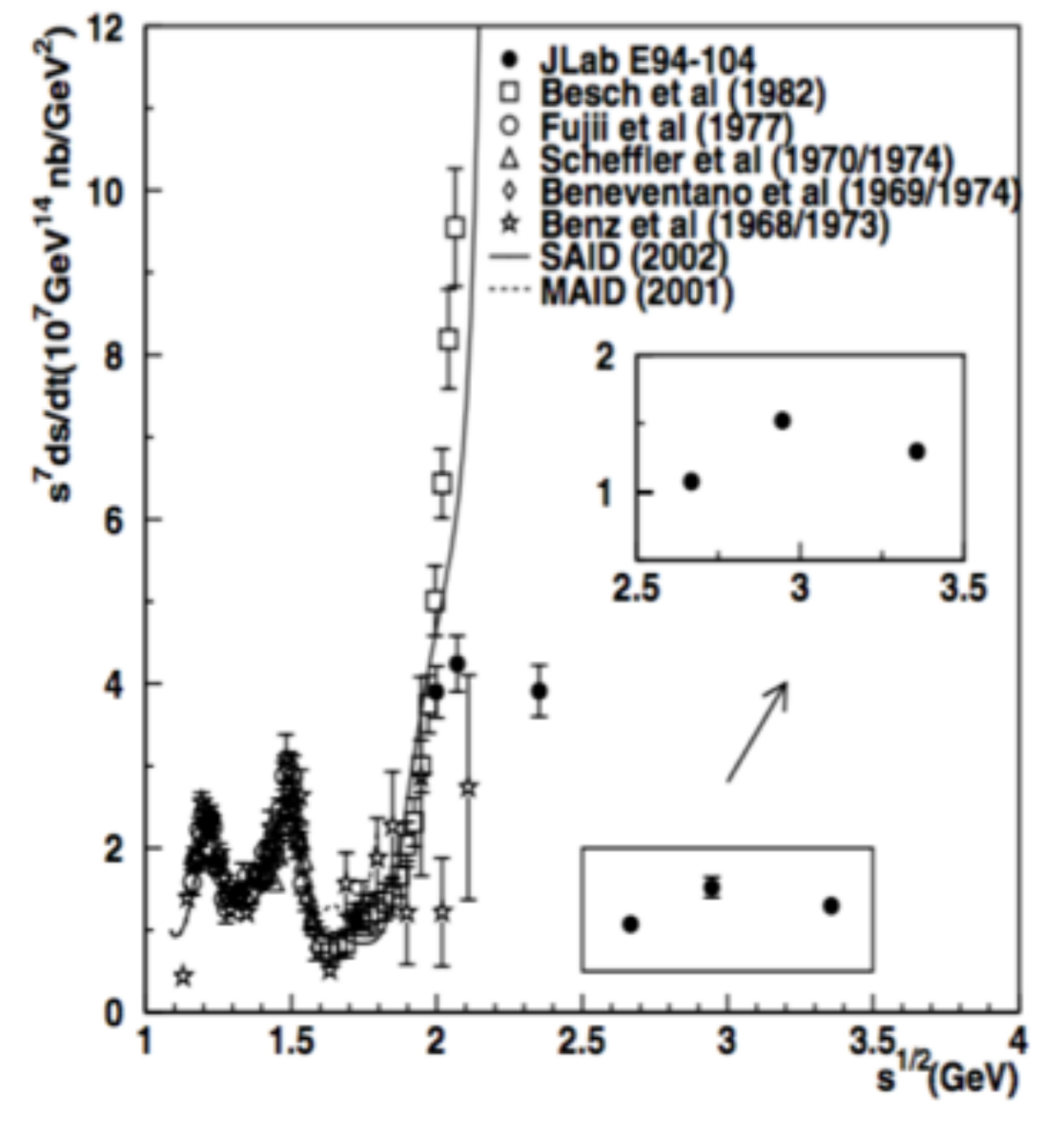}
\includegraphics[width=0.44\textwidth]{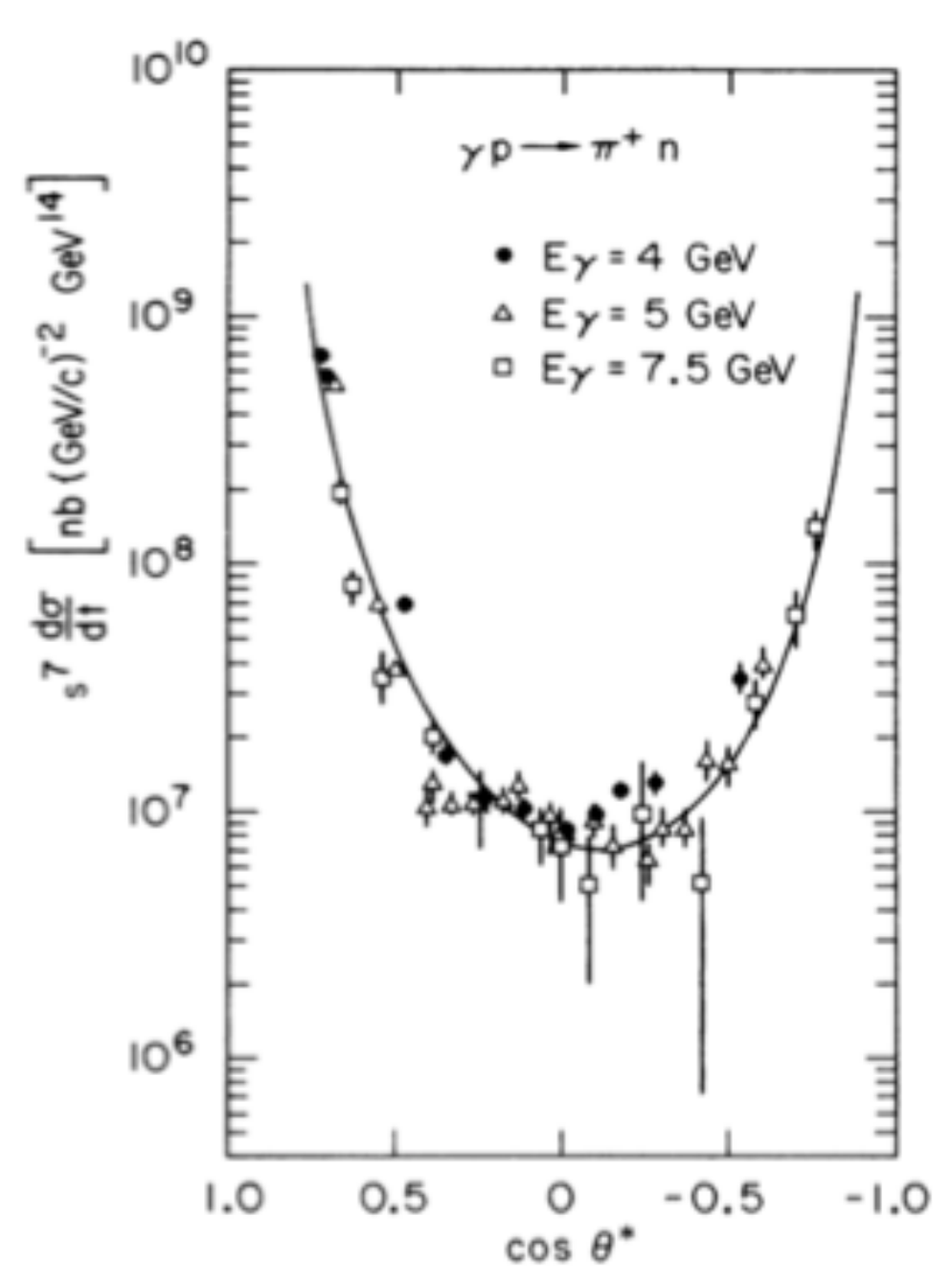}
\caption{\label{fig:photonuclear_cs}
The $s$ dependence of the photonuclear cross-section at $90^{\circ}$ in the c.m. (left) and its dependence on the c.m. angle (right). 
We extract the cross-section by fitting the $s$ dependence at high-$s$, after the low-$s$ oscillations appear to be over. 
Figures were adapted from \cite{Anderson:1976ph}.
}
\end{figure}

The cross-section for the $\gamma + n \rightarrow \pi^- + p$ reaction was calculated based on the experimental data
for $90^\circ$ scattering in the c.m. with $s > 6.25$~GeV$^2$ assuming factorization of the $s$ and c.m. angle dependence, i.e., 
$\frac{d\sigma}{dt}|_{\theta_{c.m.}} = (C\times s^{-7}) \times f(\theta_{c.m.})$, where $C$ is a free fit parameter and 
$f$ was extracted from the SLAC data assuming $f(90^\circ)=1$ \cite{Anderson:1976ph}, see 
Fig.\ \ref{fig:photonuclear_cs}. 

\begin{figure}[htpb]
\centering
\includegraphics[width=0.9\textwidth]{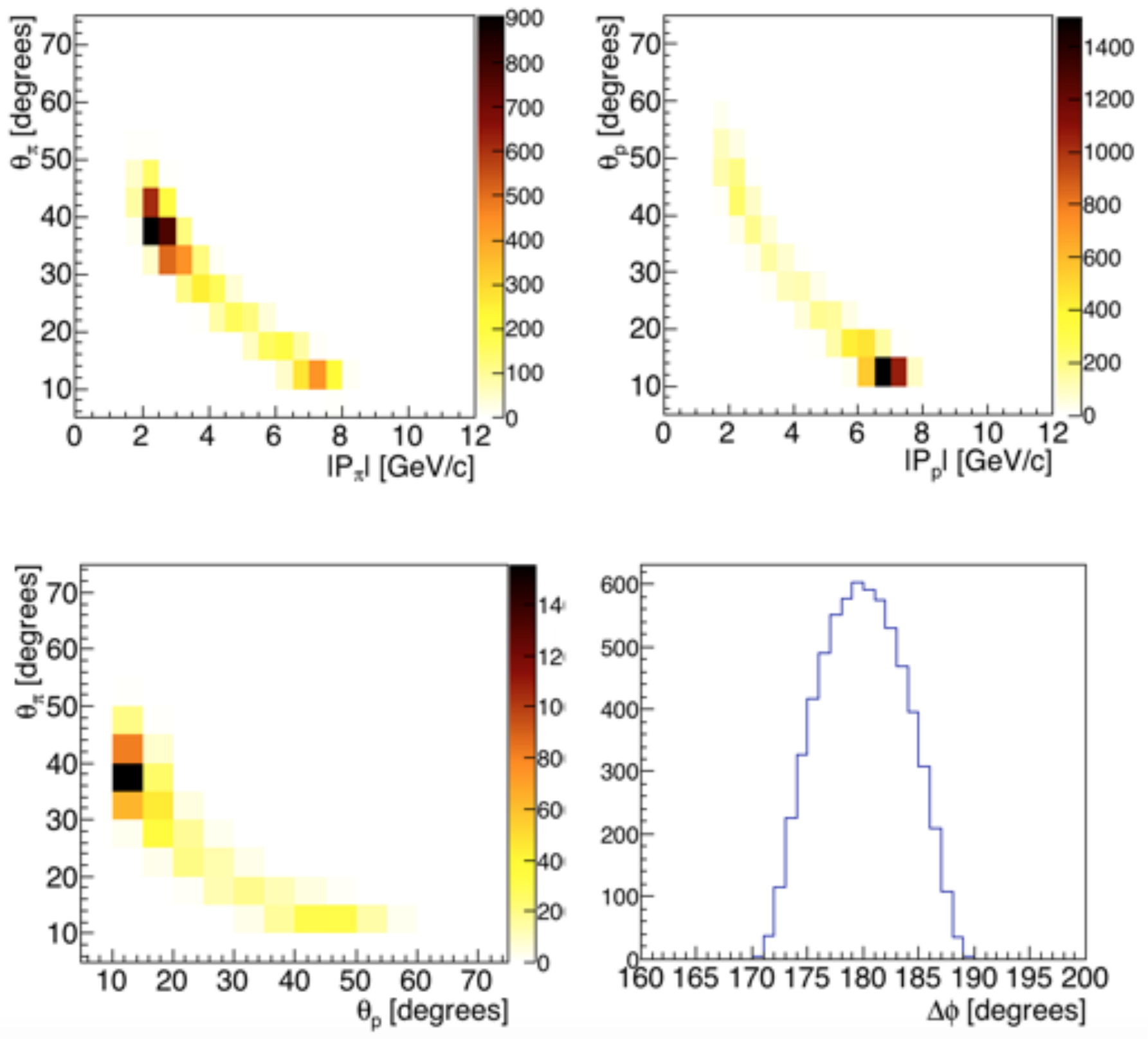}
\caption{\label{fig:kin_dists_mf}
Kinematical distributions for the final state particles of the $\gamma + n \rightarrow \pi^- + p$ 
reaction for the MF regime ($P_\text{miss} < 0.25$~GeV$/c$). 
}
\end{figure}

\begin{figure}[htpb]
\centering
\includegraphics[width=0.9\textwidth]{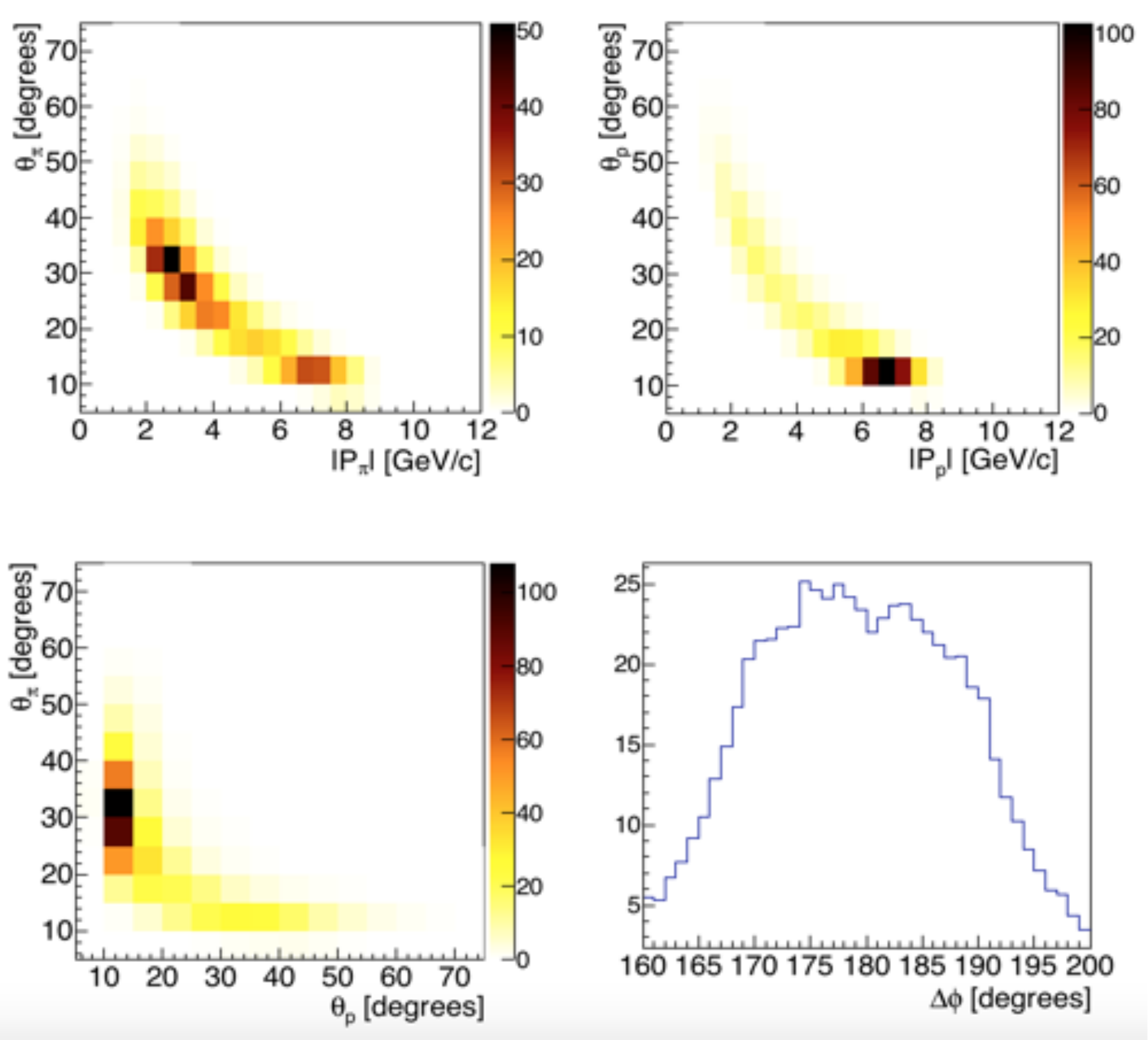}
\caption{\label{fig:kin_dists_src}
Same as Fig.\ \ref{fig:kin_dists_mf} for the SRC regime ($P_\text{miss} > 0.25$~GeV$/c$ and 
$q_\text{recoil} < 160^\circ$). 
}
\end{figure}

\begin{figure}[htpb]
\centering
\includegraphics[width=0.5\textwidth]{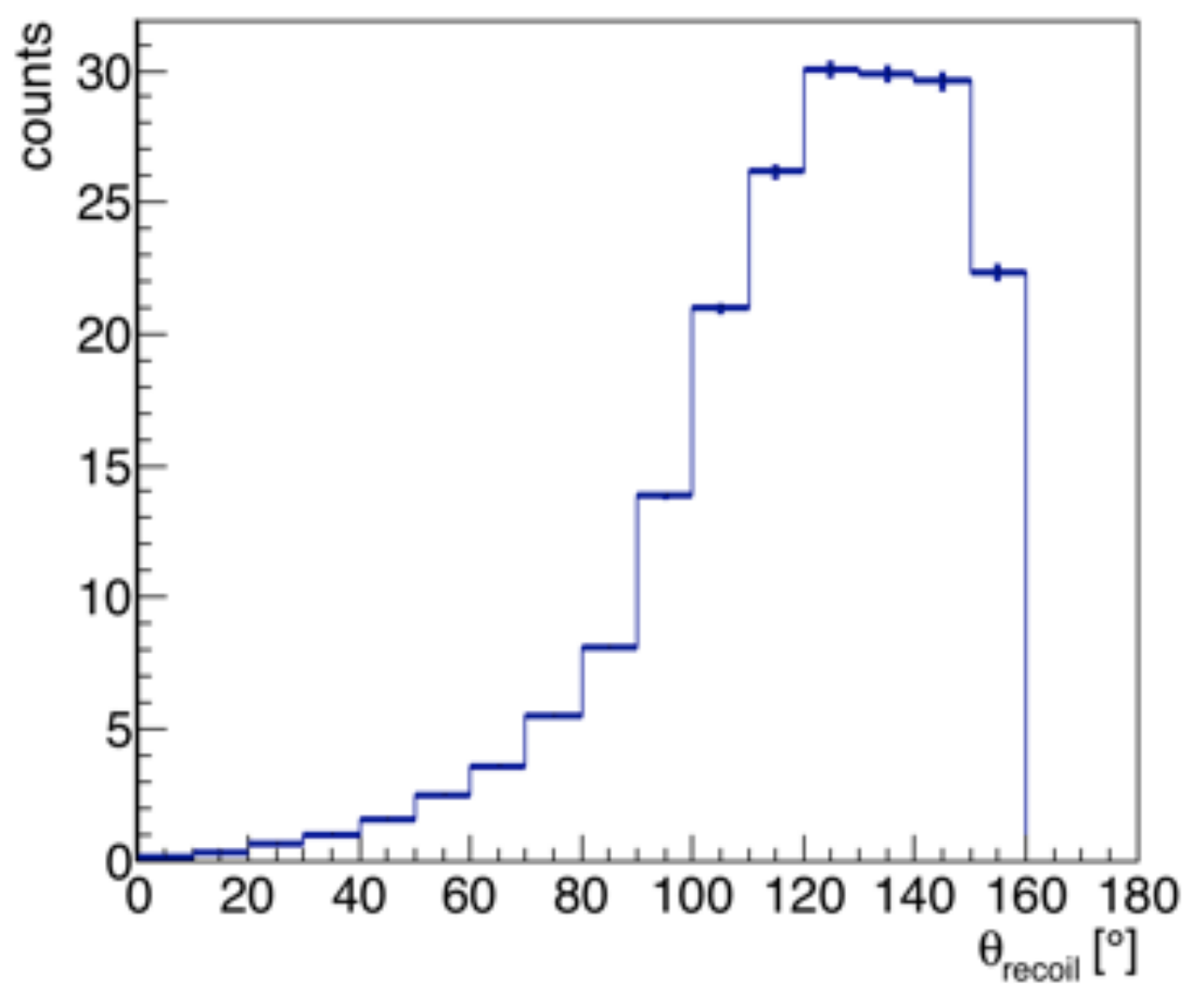}
\caption{\label{fig:theta_recoil}
The angular distribution of the recoil nucleon when scattering off an SRC pair in the nucleus. 
}
\end{figure}

The scattering was performed in the c.m. frame of the bound nucleon and gamma beam for scattering angles of 40$^\circ$--$140^\circ$. 
Hard reaction kinematics where enforced by requiring $|t,u| > 2$~GeV$^2$. We note that the rate for lower momentum transfers is very
high and within the GlueX acceptance. Figures \ref{fig:kin_dists_mf} and \ref{fig:kin_dists_src} show the kinematical distributions
for the final state particles respectively for interactions of the gamma with a mean-field nucleon and an SRC nucleon. For the case 
of SRC pair breakup, the distribution of the correlated recoil proton is shown in Fig.\ \ref{fig:theta_recoil}. As mentioned, the 
backward peak of the recoil proton is due to the $s^{-7}$ weighting of the cross-section that prefers interactions with forward
going nucleons which, in the case of SRCs, enforce the recoil nucleon to be emitted in a backward direction. We note that the resulting
kinematical distributions are not very different from those obtained for scattering off stationary nucleons, which is what GlueX was 
designed to do.

The simulation results were compared to a simple, back-of-the-envelope calculation for the reaction
cross-section. This calculation is explained in Appendix \ref{sec:cs}. The back-of-the-envelope result
is within 20\% of the simulation, giving us confidence in the validity of our simulation.

\subsection{Optimization of the Tagged Gamma Energy Range}

\label{ssec:optimization}

\begin{figure}[htpb]
\centering
\includegraphics[width=0.9\textwidth]{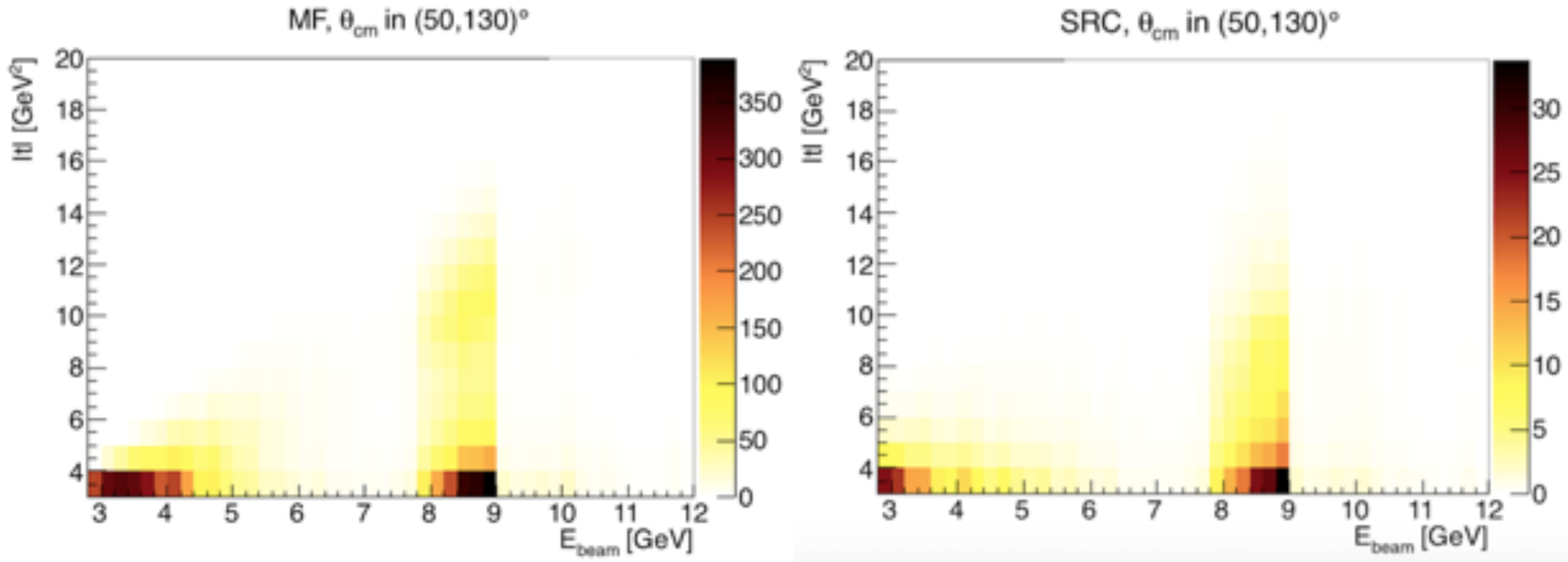}
\caption{\label{fig:gluex_t}
The momentum transfer \abs{t}  as a function of the photon beam energies for two regimes of event selection: MF on the left
and SRC on the right. The center of mass angle is in the range between $50^\circ$ and $130^\circ$.
}
\end{figure}

The Hall-D beam allows for a broad distribution of tagged photons on the target (Fig.\ \ref{fig:gluex_e}). Due to the 
coincidental rate limitation of the Hall-D tagger we cannot consider the full gamma spectrum and should focus on
a given energy range. Fig.\ \ref{fig:gluex_t} shows the correlation of $|t|$ and the beam-energy. As we are largely
interested in large $|t|$ reactions, we choose to focus the tagger at the coherent peak, with gamma energies of 
$8$~GeV~$< E_\text{beam} < 9$~GeV. 

The number of photons in the coherent peak is regulated by the size of the collimator: a wider collimator ensures more
coherent photons to hit the target. A collimator diameter of 5~mm was found to be optimal, as it allows measuring all
the coherent photons with minimal `background' from low energy photons. Smaller collimator will reduce the high-energy
gamma flux and larger collimator will increase the low energy background (leading to larger EM and neutron backgrounds)
without improving the high-energy gamma flux.

The factors limiting the beam luminosity are the coincidental rate in the tagger and the electromagnetic background 
level in the GlueX spectrometer. The tagger coincidental rate for a photon flux on the target of $2 \times 10^7$ 
photons/s and RF time of 4~ns is expected to be about 18\%.

To be conservative, the rate calculations presented below (and kinematical distribution presented above) are done for
photon beam energies in the $8$~GeV~$< E_\text{beam} < 9$~GeV range alone.

\subsection{Final State Particle Detection}

\begin{figure}[htpb]
\centering
\includegraphics[width=0.49\textwidth]{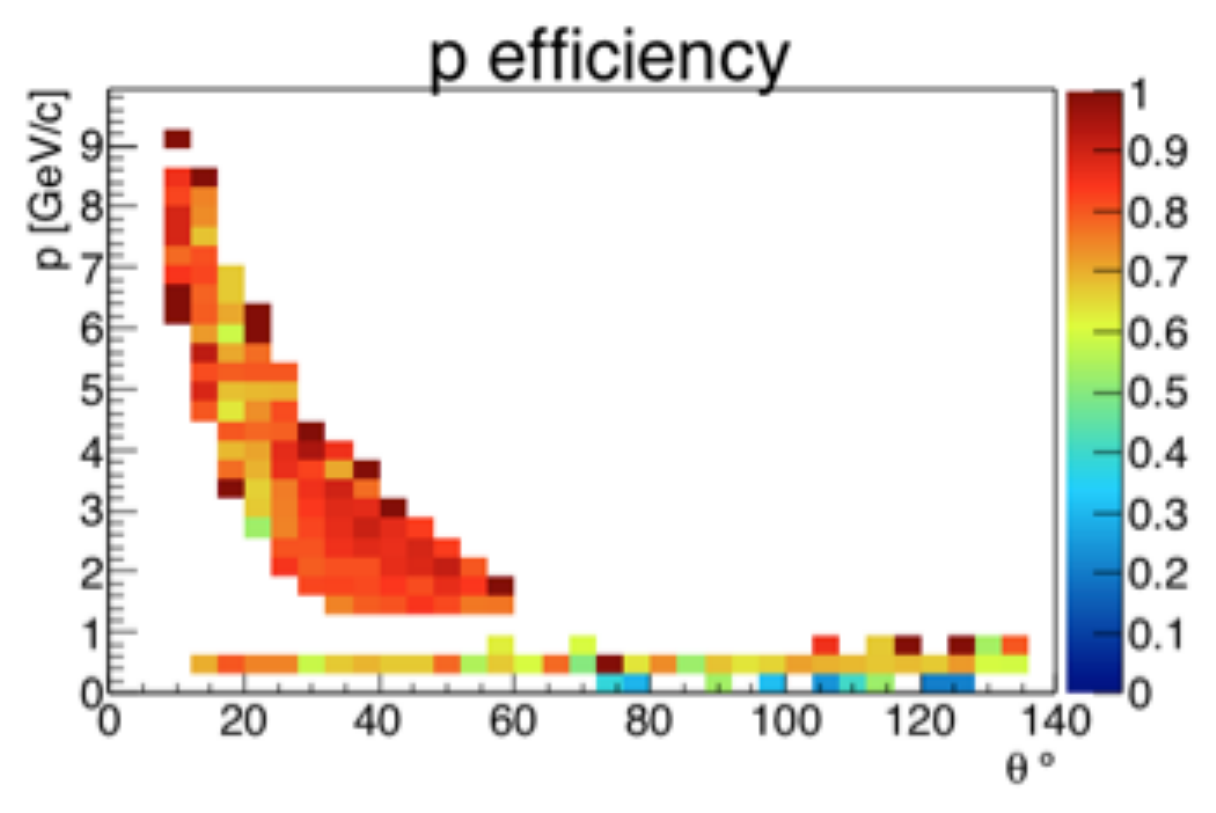}
\includegraphics[width=0.49\textwidth]{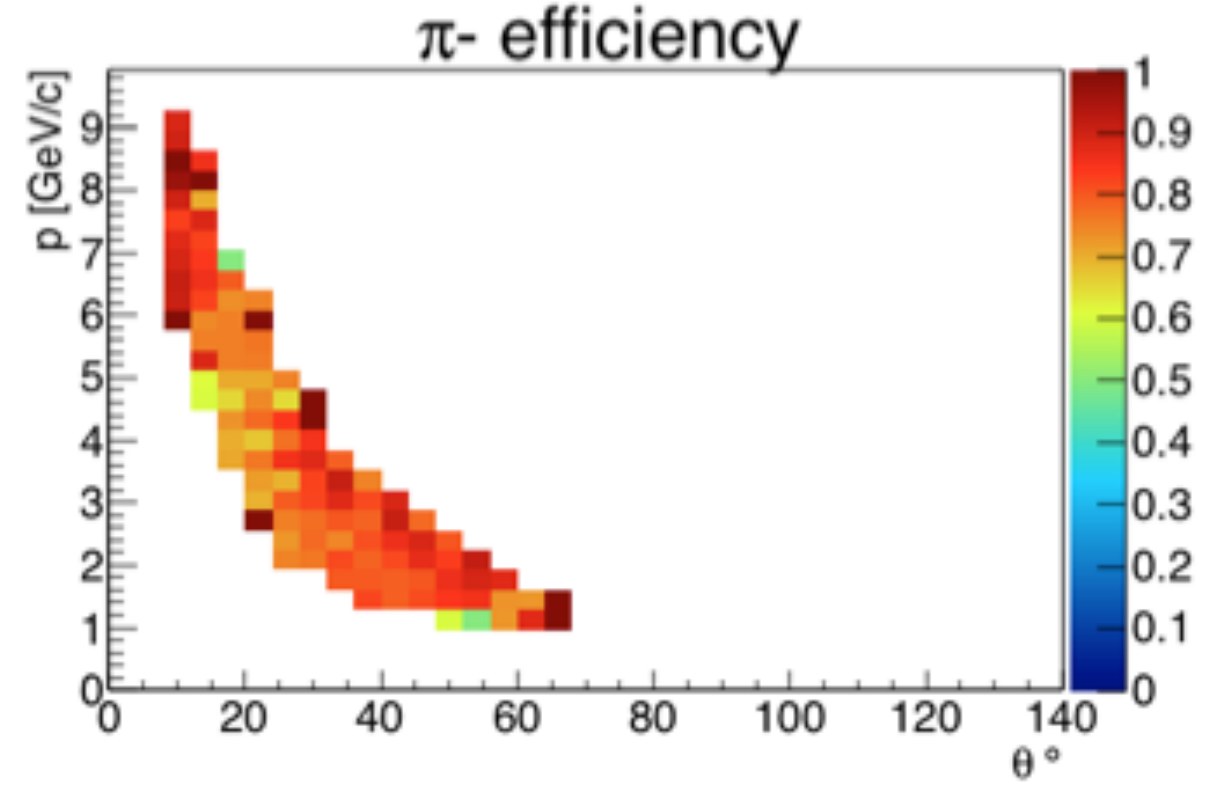}
\caption{\label{fig:eff}
The reconstruction efficiency for the charged tracks coming from the reaction $\gamma n \rightarrow \pi^- p$. 
On the left panel the low momentum band is due to recoil protons from SRC pair breakup.
}
\end{figure}

The efficiencies for the reconstruction of final state particles (i.e. meson-baryon pair and, in the case of SRC breakup,
also the recoil proton) were simulated using the Geant-based GlueX simulation chain for the event generator described in
the previous section. Fig.\ \ref{fig:eff} shows the simulated detection efficiency for each particle separately. The average 
efficiency for the simultaneous reconstruction of the proton and a pion was found to equal 64\%.

Based on the current GlueX data reconstruction efficiencies, we expect that more complex final states will have 
varying detection efficiencies reaching down to 30\% for rho mesons. The total detection efficiency for the reaction
$\gamma + n \rightarrow \rho^- + p$ is therefore assumed to be $0.9 \times 0.3 = 27\%$.

\begin{figure}[htpb]
\centering
\includegraphics[width=0.9\textwidth]{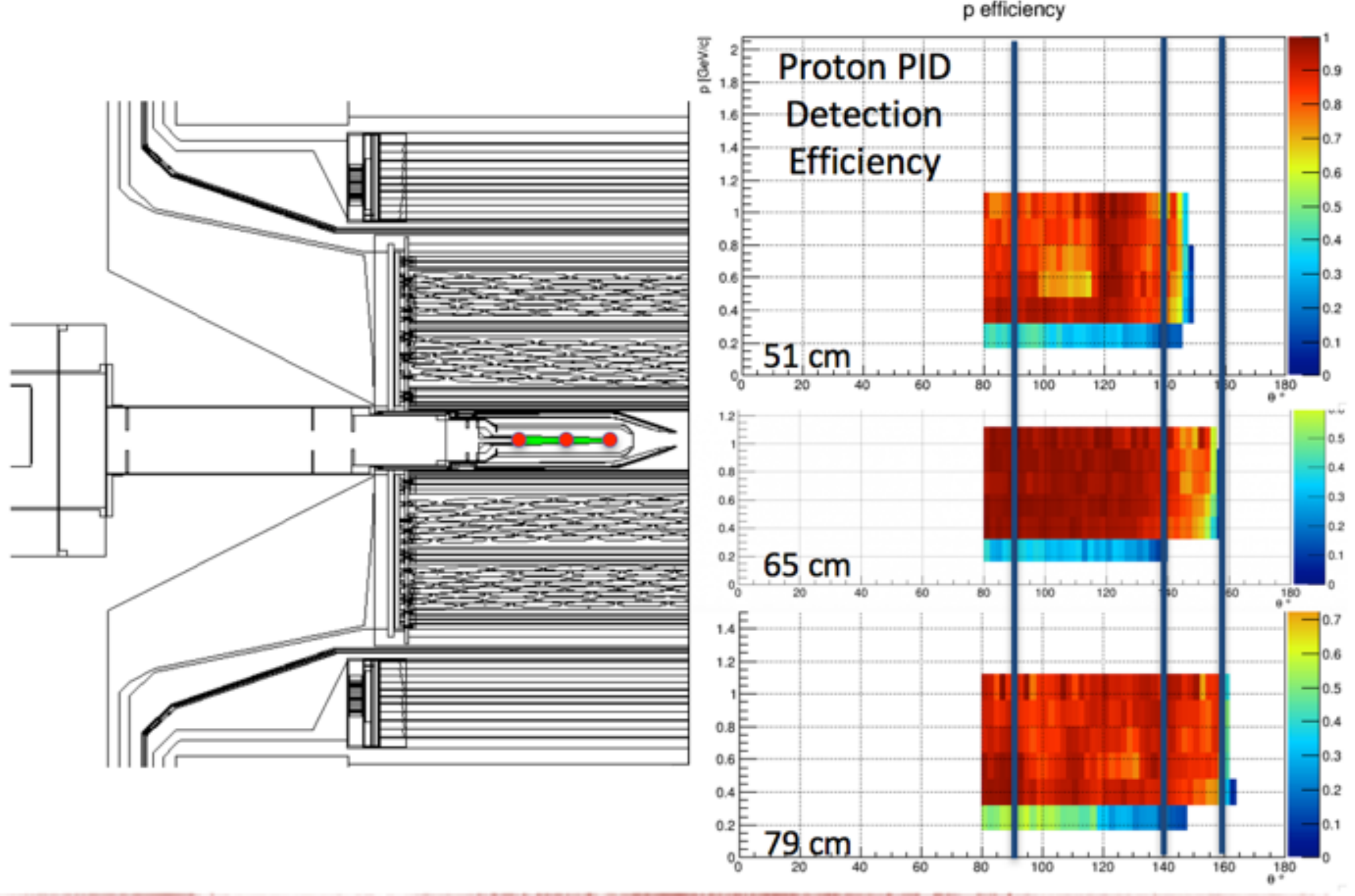}
\caption{\label{fig:eff_p_recoil}
The detection efficiency for recoiling protons in GlueX as a function of the recoil angle and momentum for 3 different vertex locations.
}
\end{figure}

\begin{figure}[htpb]
\centering
\includegraphics[width=0.6\textwidth]{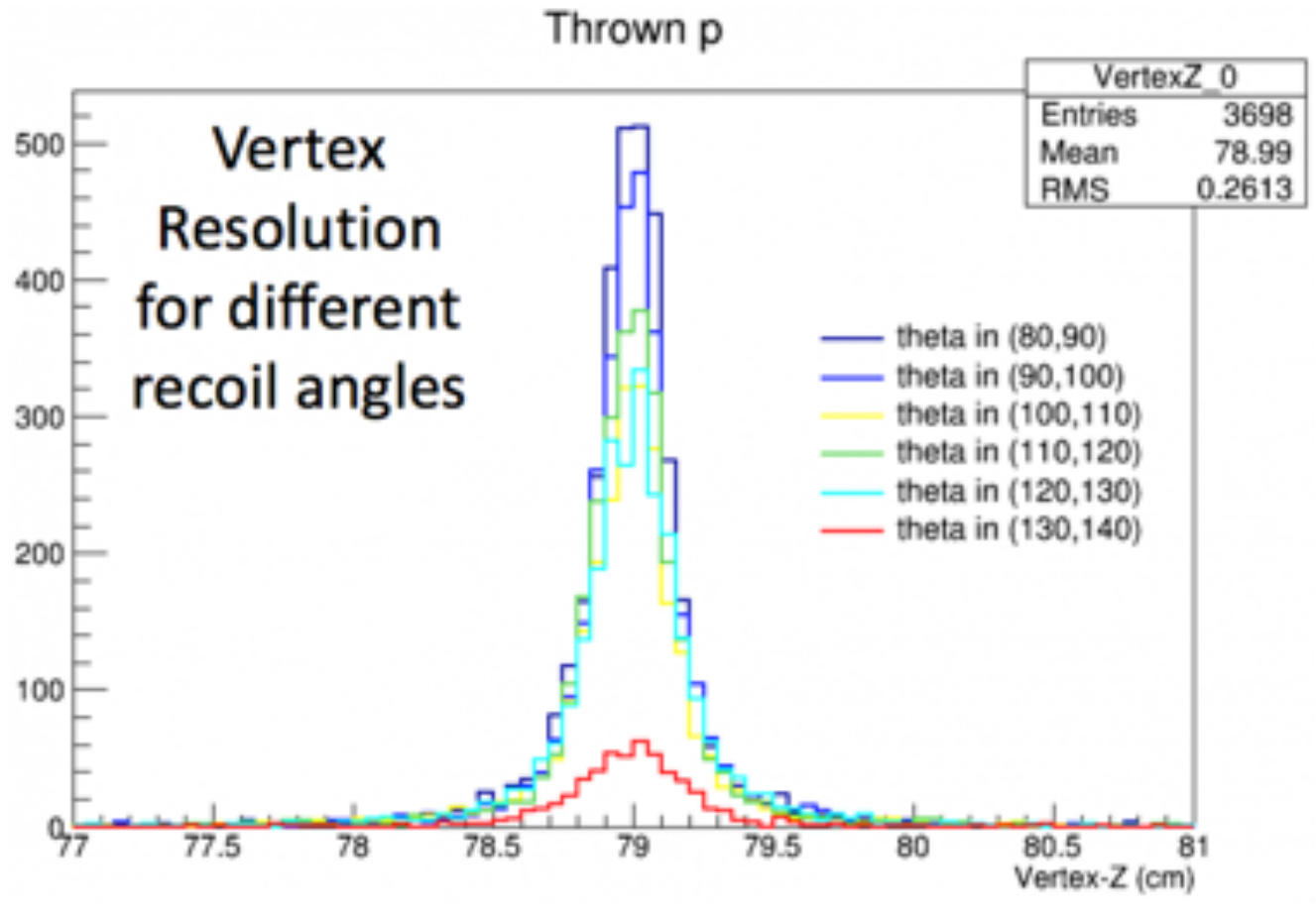}
\caption{\label{fig:res_vertex}
The vertex reconstruction resolution for recoiling protons at various recoil angles.
}
\end{figure}

In the case of SRC pair breakup, one should also take into account the detection efficiency of the recoil proton
in the backward hemisphere. Figure \ref{fig:eff} also shows the simulated detection efficiency of the recoil proton
for the case of SRC pair breakup. Fig.\ \ref{fig:eff_p_recoil} shows the detailed acceptance and detection efficiency
for recoiling protons for 3 different vertex positions---at the center and two ends of the 30~cm long hydrogen target. 
As can be seen, the detection efficiency is very high up to $140^\circ$ in the lab and can extend up to $160^\circ$ 
when the target is placed downstream. Fig.\ \ref{fig:res_vertex} shows the vertex reconstruction resolution, showing 
we can separate solid target foils with a distance of $\sim 1$~cm which is more than enough to separate different 
foils in the case of $^{12}$C.

\subsection{Expected Rates}

The rate calculations were done for the $\gamma + n \rightarrow \pi^- + p$ and $\gamma + n \rightarrow \rho^- + p$ reactions, 
using the simulation presented in section \ref{ssec:kinematics}. We choose these two reactions as they have the 
smallest and largest cross-sections respectively, of the reactions listed in table \ref{table:reactions} which makes them a good 
representative of the various expected rates.

We assume a total of 30~beam days with a photon flux of $2\times 10^7$~s$^{-1}$ (compared to the nominal GlueX photon 
flux of $10^8$~photons/s) and four targets: D, $^{4}$He, and $^{12}$C. Based on the acceptance 
simulations presented above, we assume 80\% detection efficiency for each of the leading baryon and meson and 65\% 
for the recoil nucleon. We assume the nominal nuclear attenuation effect reduces the total cross-section as $A^{-1/3}$. 
Table \ref{table:targets} lists the parameters for the chosen targets. The factors limiting the event rates are the following:
\begin{itemize}
\item GlueX detector capabilities (maximum possible gamma flux on target)
\item Electromagnetic background in the GlueX spectrometer
\item Tagger coincidental rate
\item Neutron background
\end{itemize}
Presently, GlueX is operating with a 30~cm long liquid hydrogen (LH) target (3.4\%~radiation length, $X_0$). For the nominal
beam flux on the target of $10^8$~photons/s the electromagnetic background is reaching its upper limit. In order
to comply with the electromagnetic background limits, we assume the carbon target thickness to be 7\% $X_0$ (note that 
unlike for hydrogen, for nuclei there are 2 nucleons for each electron in the target. Therefore, for EM background
estimations, 7\% $X_0$ on nuclei is equivalent to $3.5\%$~$X_0$ on hydrogen). The radiation lengths for liquid hydrogen, 
deuterium, and helium are similar. The use of nuclear targets will increase the slow neutron background that can induce 
some damage to GlueX detector components such as the SiPMs. Table \ref{table:targets} shows an estimate of the neutron 
background, done in collaboration with Hall D staff, based on JLAB-TN-11-005. GlueX was designed to handle one year of 
LH running with a photon flux of $10^8$ photons/s. While the proposed gamma flux for this measurement is 
smaller by a factor of 5, detailed estimations by the Radcon group, backed with relevant measurements, show that replacing 
the LH target with $^4$He will increase the neutron background by a factor of 4--5 (depending on the exact location). 
This implies that the neutron background for $^4$He target in our running conditions will be similar to the GlueX design 
specifications. The estimated neutron backgrounds for all targets are shown in table \ref{table:targets}, and are calculated
from the Radcon $^4$He estimate and their reported $A$-dependence. For reference, the table includes the GlueX LH target
under nominal running conditions, i.e. $10^8$ photons/s. The background rates for the other targets take into account the
five-fold reduced gamma flux for the proposed experiment. While this background estimation procedure takes into account the 
main differences coming from nuclear targets relative to LH, the deuteron backgrounds could be somewhat higher. 
Given the very short deuteron beam time (table \ref{table:rates}) this should not be an issue with regard to integrated damage.

\begin{table}
\centering
\caption{\label{table:targets}
Parameters for the proposed targets. The current GlueX liquid hydrogen target (LH) is shown for comparison.
(*) The neutron flux for the LH target is taken under assumption of the nominal flux of $10^8$ photons/s in 
the coherent peak. 
}
\vspace{5pt}
\begin{tabular}{| c | c | c | c | c | }
\hline \T \B
{Target} & \shortstack[c]{Thickness \\ $[\text{cm}]$ / \% $X_0$} & \shortstack[c]{Atoms/cm$^{2}$ \\ for the given target \\ thickness} & \shortstack[c]{EM bkg. rel. \\ to GlueX} 
& \shortstack[c]{Neutron bkg. \\ rel. to GlueX}  \T \B \\ 
\hline
D & 30 / 4.1 & $1.51 \times 10^{24}$ & 0.5 & 1.3 \T \B \\
\hline
$^4$He & 30 / 4 & $5.68 \times 10^{23}$ & 0.5 & 1 \T \B \\
\hline
$^{12}$C & 1.9 / 7 & $1.45 \times 10^{23}$ & 1 & 0.8 \T \B \\
\hline
%$^{40}$Ca & 0.73 / 7 & $1.70 \times 10^{22}$ & 1 & 0.3 \T \B \\
%\hline
LH & 30 / 3.4 & $1.28 \times 10^{24}$ & 1 & 1*  \T \B \\
\hline
\end{tabular}
\end{table}
 
\begin{figure}[htpb]
\centering
\includegraphics[width=0.49\textwidth]{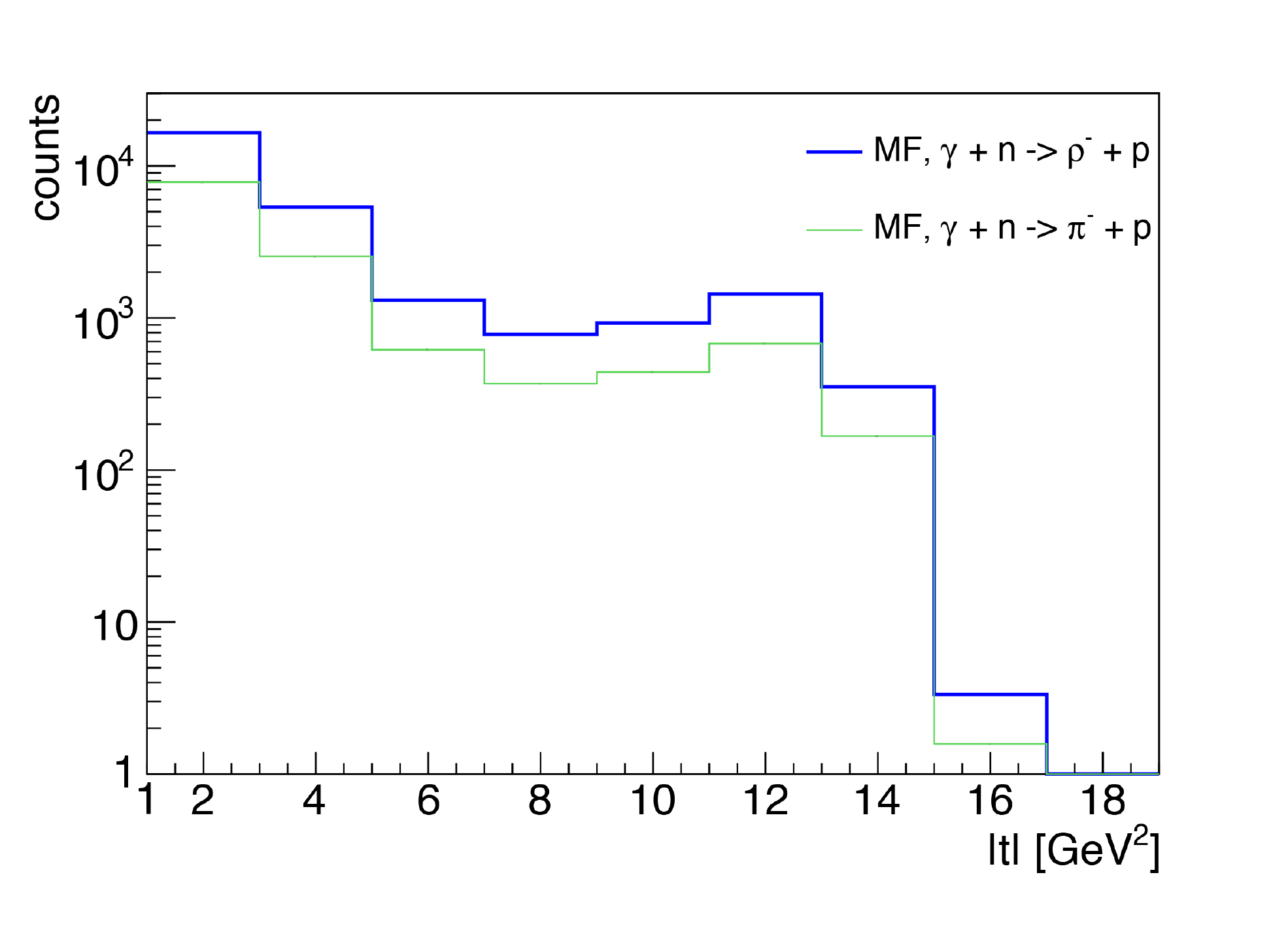}
\includegraphics[width=0.47\textwidth]{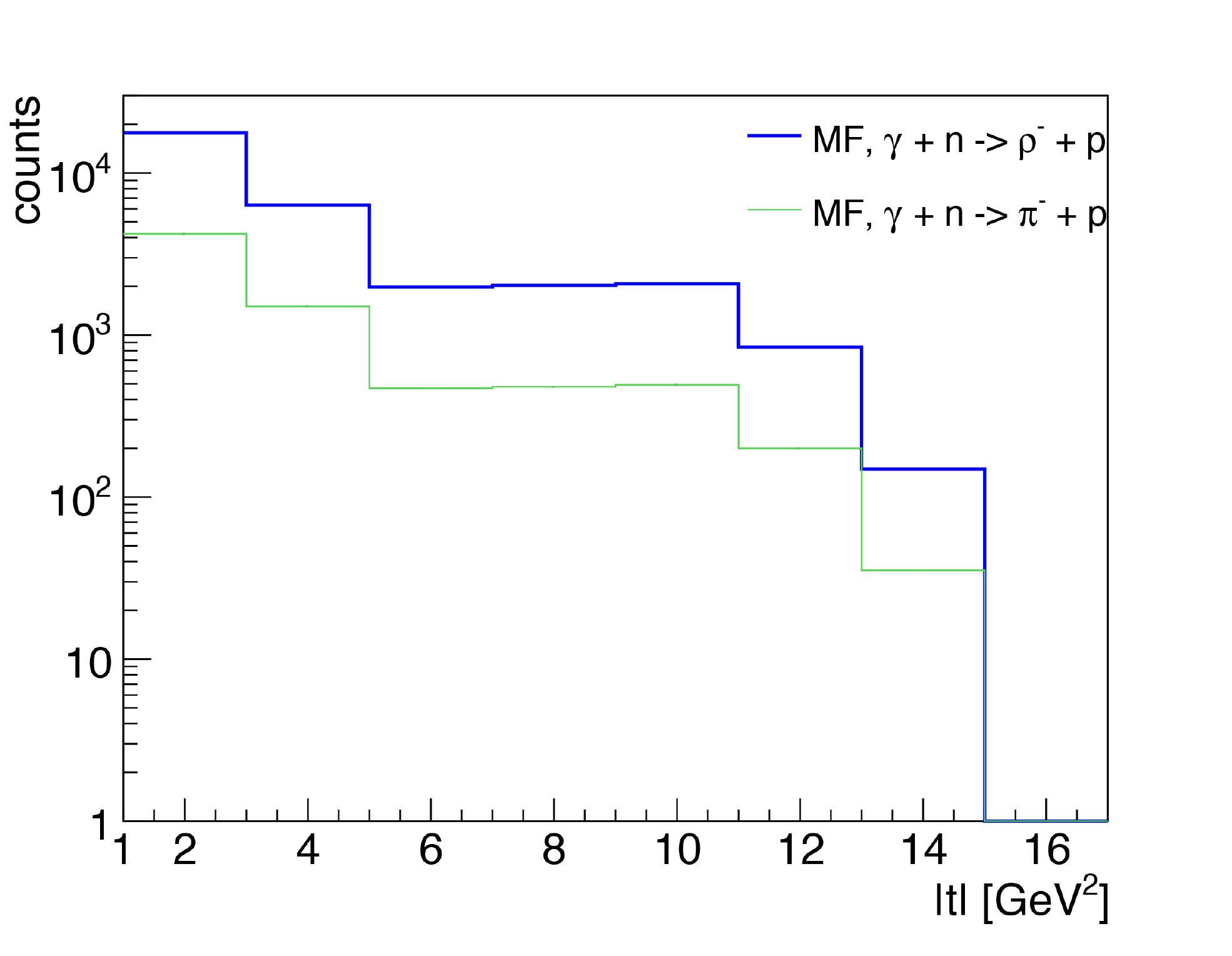}
\caption{\label{fig:rates_mf}
The expected count rate for 10 days running as a function of $|t|$ for Deuterium (left) and $^{12}$C (right) 
targets in mean-field kinematics for two different reactions.
}
\end{figure}

Table \ref{table:rates} lists the expected events rate and beam time for each target for $|t,u|>2$~GeV$^2$ 
for mean-field and SRC events separately. For Deuterium, as we use the AV18 distribution, the distinction 
is based on the initial momentum of the nucleon that the gamma interacted with (above or below 250~MeV$/c$). 
Figure \ref{fig:rates_mf} shows the expected count rate of various $|t|$ bins for Deuterium and $^{12}$C for mean-field 
events. The statistics for $|t| < 2$ (GeV/c)$^2$ is rich, allowing to map the transition between different transparency regimes. 
Other nuclei have the same $|t|$ dependence and the expected count rate per bin can be scaled based 
on the total number of events listed in Table \ref{table:rates}. We have chosen to distribute the beam time
between the different targets so as to obtain comparable discriminating power for transparency 
studies and scaling of SRC pairs; the larger nuclei have larger predicted effects and therefore fewer 
statistics are needed to observe them at similar levels of significance.

\begin{table}
\centering
\caption{\label{table:rates}
Event rates estimation. See text for details.
}
\vspace{5pt}
\begin{tabular}{ | l | l | l | l | l | l | }
\hline
\multirow{2}{*}{Target} & \multicolumn{2}{l |}{$\gamma + n \rightarrow \pi^- p$} & \multicolumn{2}{l |}{$\gamma + n \rightarrow \rho^- p$} & 
\multirow{2}{*}{\shortstack[c]{PAC \\ Days}} \\ \cline{2-5} \T \B
& MF & SRC & MF & SRC & \T \B \\
\hline
D &          13,600 &   750 & 57,000 &  3,000 & 5 \T \B \\
\hline
%%$^4$He &     13,000 & 670 & 54,500 & 2,800 & 8 \T \B \\
$^4$He &     16,000 & 840 & 68,000 & 3,500 & 10 \T \B \\
\hline
%%$^{12}$C &    7,400 & 2,300 & 31,000 & 9,500 & 10 \T \B \\
$^{12}$C &    8,900 & 2,800 & 37,000 & 11,000 & 12 \T \B \\
%\hline
%$^{40}$Ca &   2,600 &   840 & 10,900 &  3,500 & 14 \T \B \\
\hline
\multicolumn{5}{|r|}{Calibration, commissioning, and overhead:} & 3 \T \B \\
\hline
\multicolumn{5}{|r|}{\textbf{Total PAC Days:}} & 30 \T \B \\
\hline
\end{tabular}
\end{table}

\begin{figure}[htpb]
\centering
\includegraphics[width=0.49\textwidth]{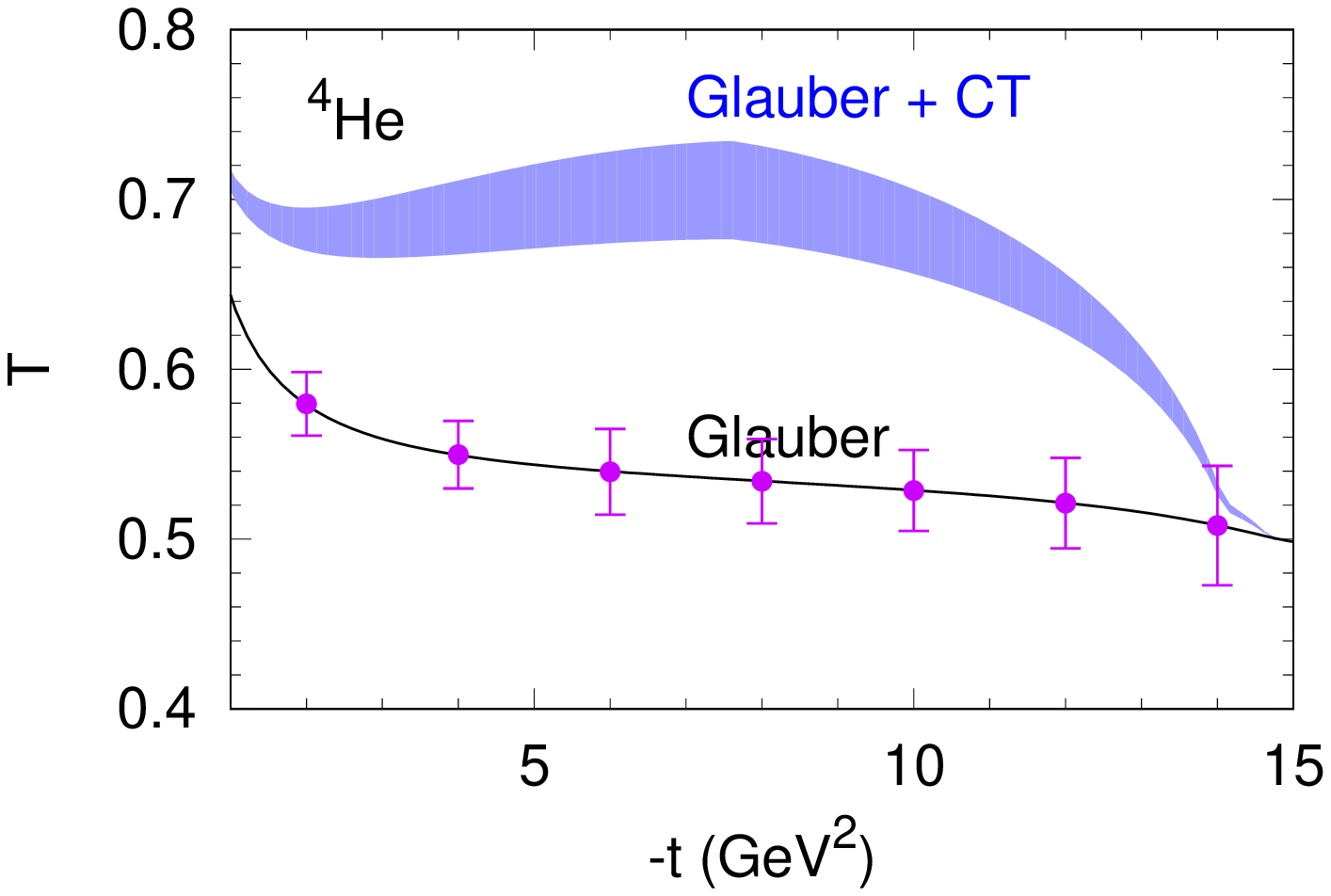}
\includegraphics[width=0.49\textwidth]{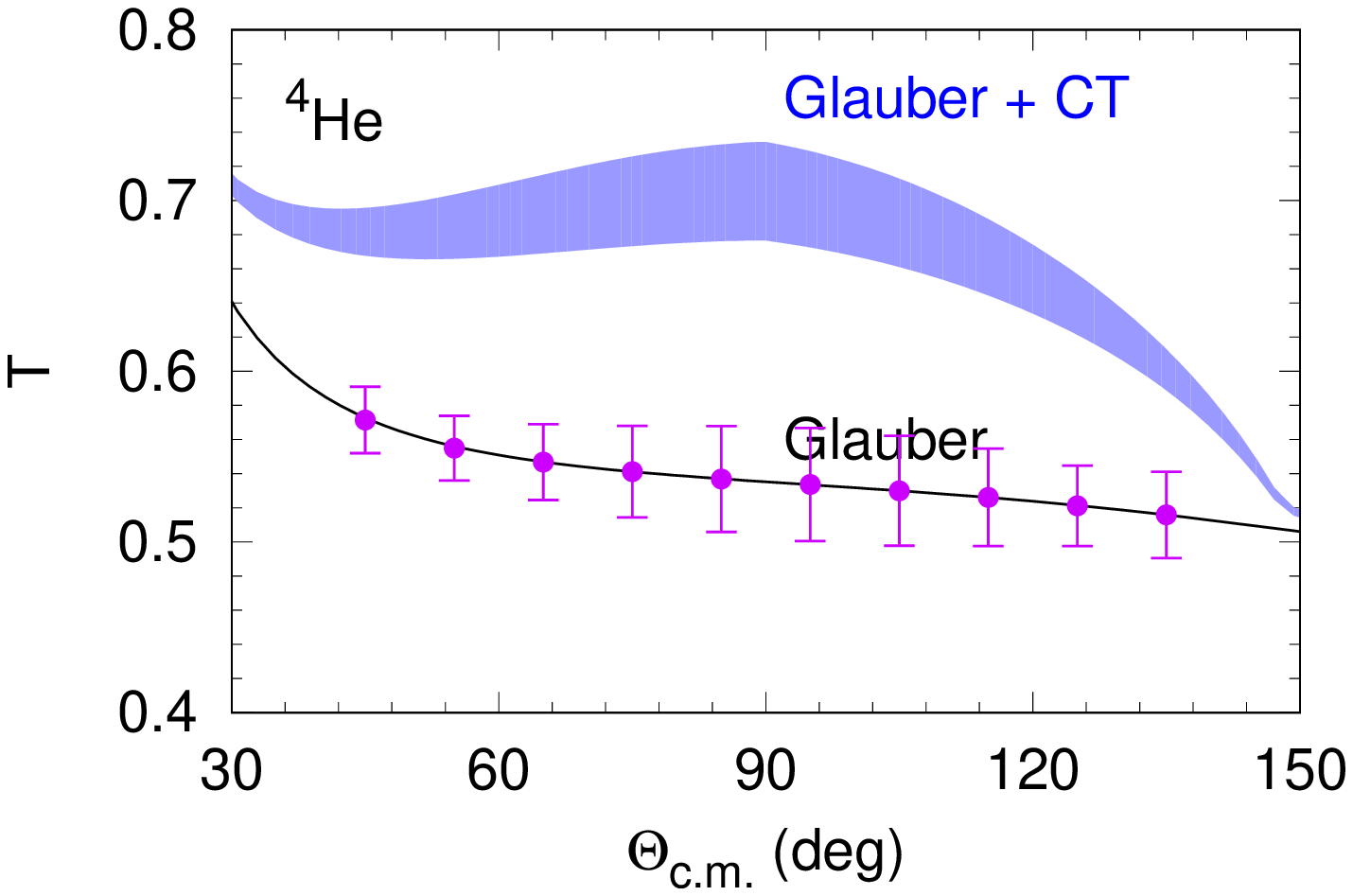}\\
\includegraphics[width=0.49\textwidth]{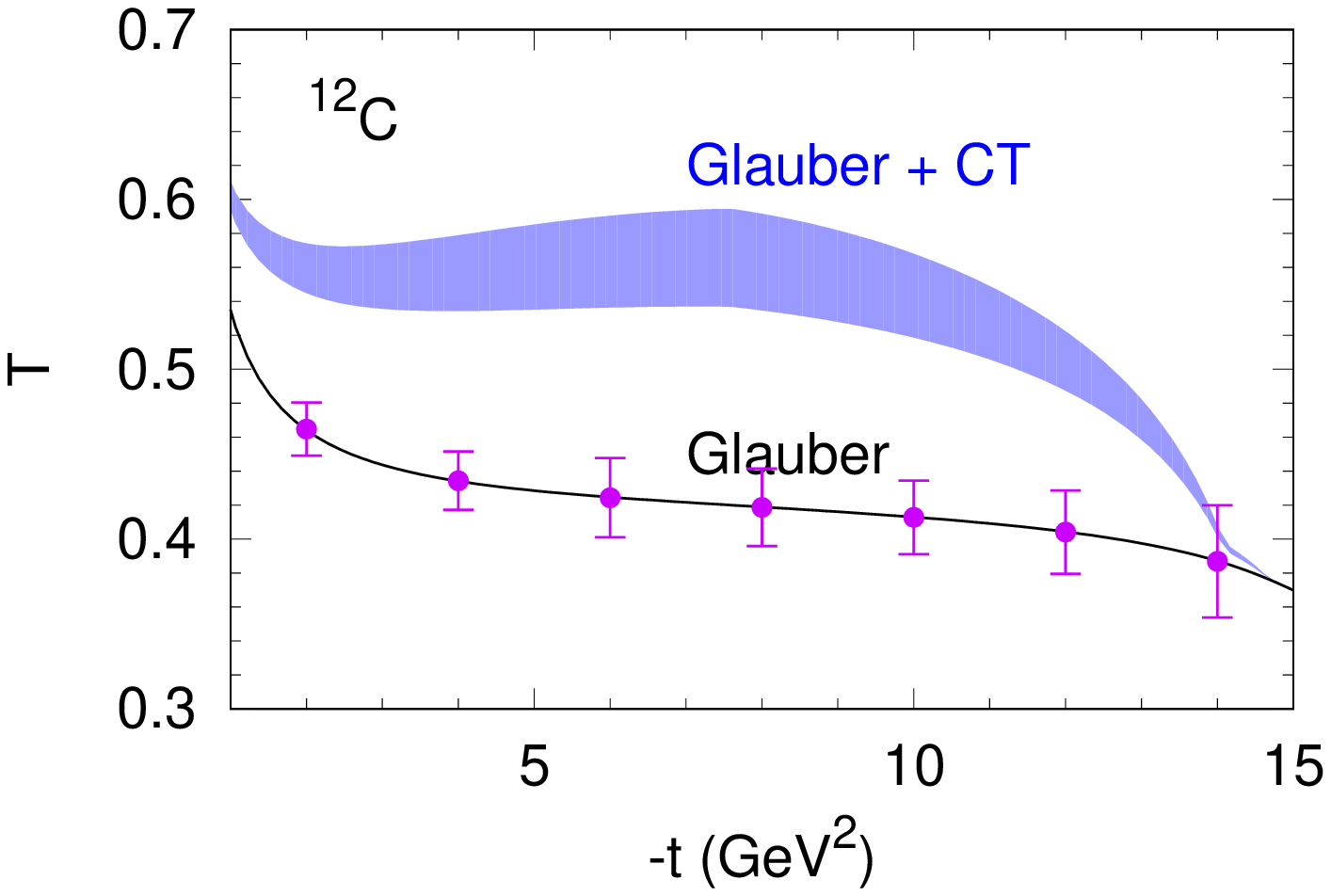}
\includegraphics[width=0.49\textwidth]{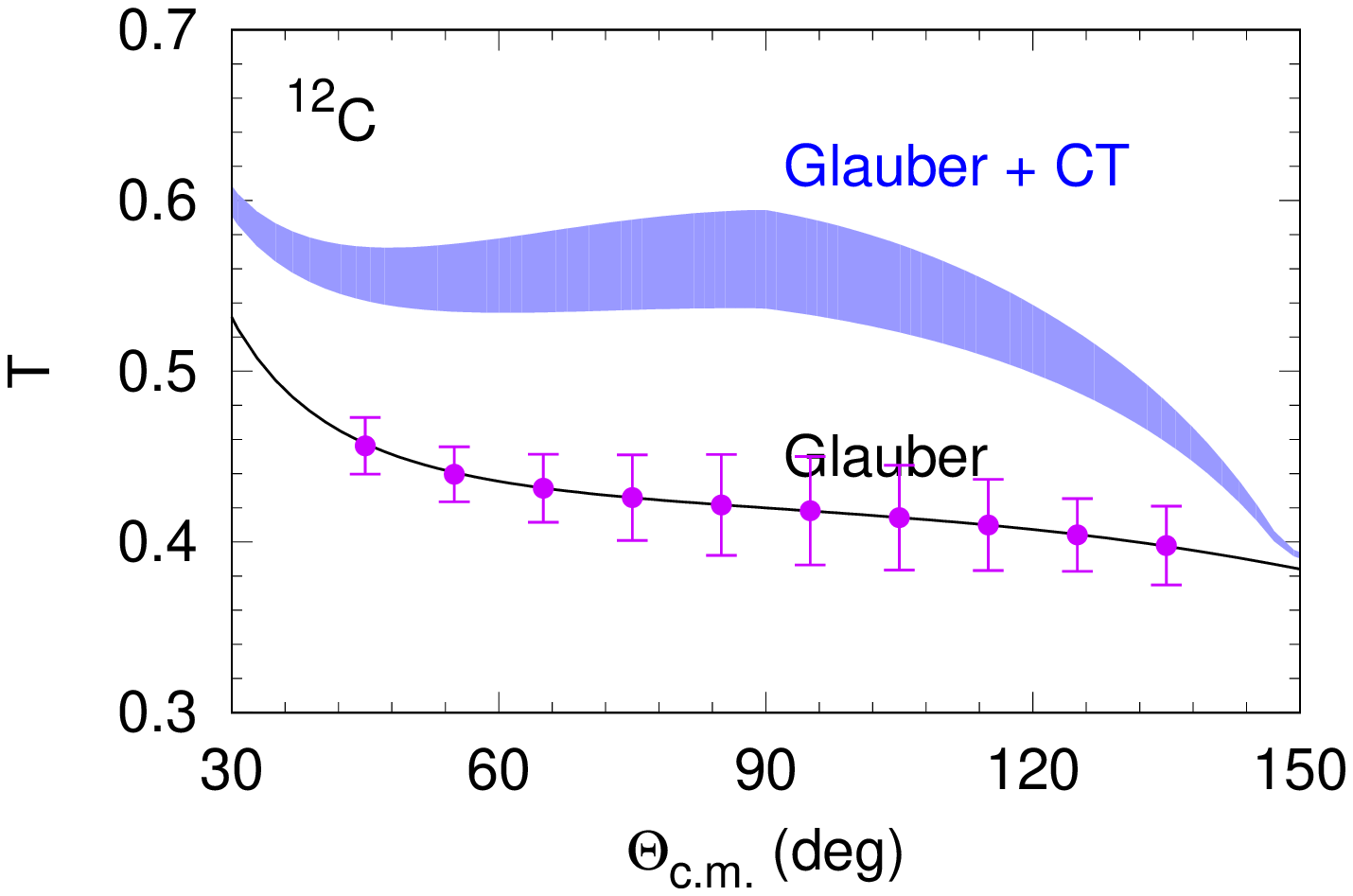}\\
\caption{\label{fig:rates_trans}
Expected uncertainties (statistical + systematical) for the measurement of the $\gamma + n \rightarrow \pi^- + p$ reaction 
off $^{4}$He (upper row), and $^{12}$C (lower row).
}
\end{figure}

Fig. \ref{fig:rates_trans} shows the expected results for the color transparency for the 
$\gamma + n \rightarrow \pi^- + p$ reaction. Other reactions from table \ref{table:reactions}
will have comparable or better discriminating power. We note that by taking ratios for nuclei relative to 
deuterium we minimize many of the theoretical systematical uncertainties and are dominated by the beam flux 
and target densities. Both are expected to be known to better than 3\% which is what we assume for the overall 
systematical uncertainty which is included in Fig.~\ref{fig:rates_trans}.

\begin{figure}[htpb]
\centering
\includegraphics[width=0.6\textwidth]{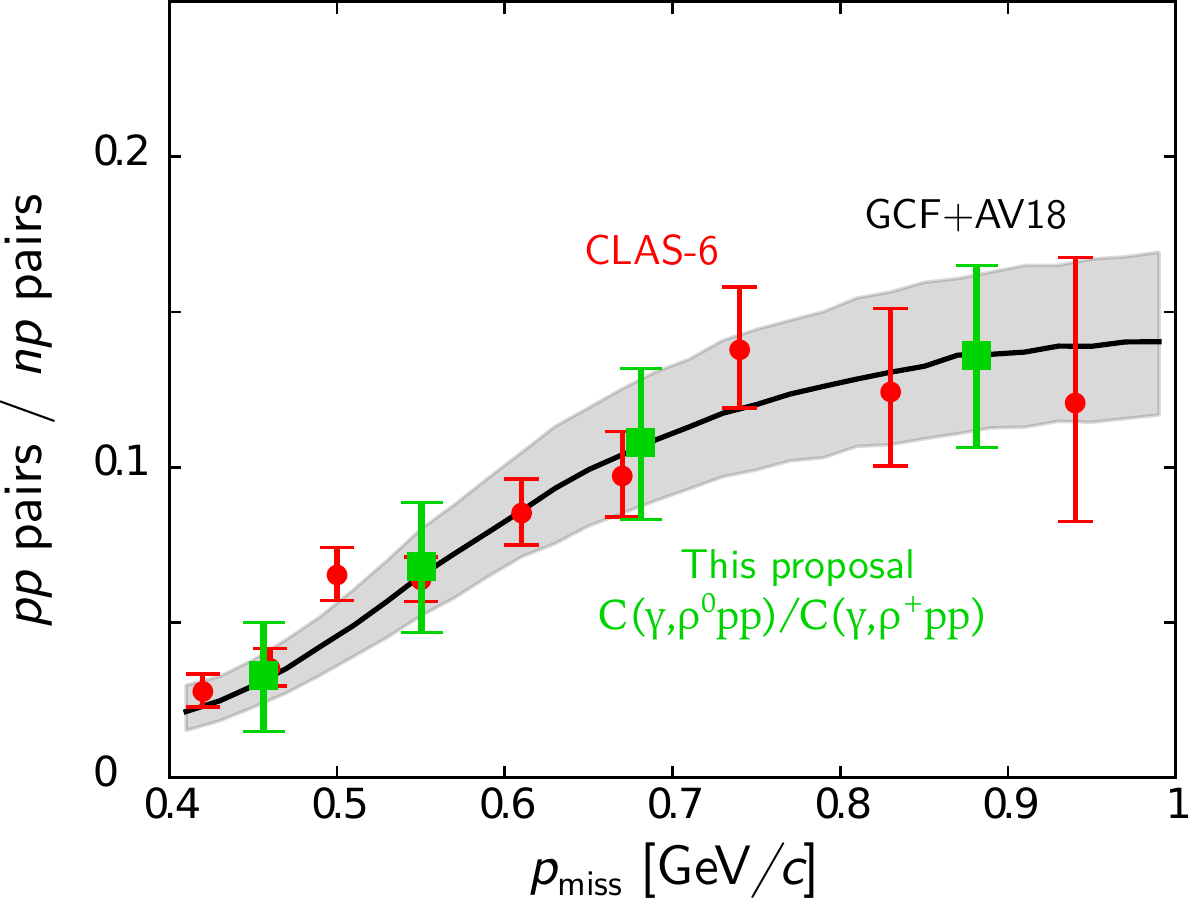}
\caption{\label{fig:reach_src}
The expected precision of this proposal for testing $np$-dominance via the $\rho$ production channel, compared
with electron scattering data from CLAS-6.
}
\end{figure}

For SRC studies, while the rates are modest, they are in fact comparable and even higher per reaction 
as compared to the 6~GeV measurements done in Hall A and B. Therefore, the cross-section ratios for 
scattering off nuclei relative to deuterium, in SRC kinematics, will allow us to test the observed 
$np$-dominance and extract the relative number of SRC pairs in the measured nuclei with $<4\%$--$10\%$
accuracy, depending on the channel considered. As an example, Fig.~\ref{fig:reach_src} shows the
expected sensitivity for $\rho$-production, the highest-rate channel, compared with the preliminary
analysis of CLAS data (shown previously in Fig.~\ref{fig:pp2p}). The precision will be sufficient
to confirm both $np$-dominance and explore the transition to the repulsive core. These data will
therefore improve our understanding of SRCs and help to reduce interpretation
uncertainty in a unique way un-matched by any other measurement that can be performed at JLab.

\section{Complementary Experiments at JLab}

The goals of our proposed measurement run complementary to those of several other approved JLab 12~GeV
experiments, which will be described in this section. 

%% Short-range correlations

Semi-inclusive and exclusive measurements of short-range correlations are the focus of several recent
and upcoming experiments, which are complimentary to our proposed SRC program. The Hall A experiment 
E12-14-011~\cite{Hen:2014gna} ran in 2018, measuring asymmetries between the mirror nuclei $^{3}$H
and $^3$He to study the isospin-dependent effects in SRCs~\cite{Cruz-Torres:2019bqw}, along with an
inclusive counterpart, E12-11-112 \cite{E12:11:112}. The Hall C experiment E12-17-005~\cite{E12:17:005}
will also look for isospin-dependent effects through measurements on $^{40}$Ca, $^{48}$Ca, and $^{54}$Fe.
The Hall B run group proposal E12-17-006 includes measurements of short range correlations in a wide 
range of targets~\cite{E12:17:006a}. 

The interpretation of the above experiments depends on the general framework for our understanding of SRCs,
FSIs, and reaction mechanisms in electron-induced pair break-up, which our proposed measurement will attempt
to validate with the new method of photo-induced reactions. 

%% Nucleon modification

Understanding of the modification of nucleon structure within the nuclear medium is a salient topic in nuclear
physics under active investigation in several JLab experiments. Our proposal focuses on a completely novel 
observable, branching ratio modification, which would complement traditional ``EMC'' electron-scattering
measurements. Among the upcoming electron scattering measurements, we highlight E12-11-003A~\cite{E12:11:003A}
and E12-11-107 \cite{Hen:2014vua}, which will test the role that highly-virtual nucleons play in the EMC
effect by looking at recoil-tagged $F_2$ structure functions of bound nucleons in deuterium. 

%% Color transparency

Color transparency for mesons, hints of which were seen in the JLab 6~GeV program \cite{Clasie:2007aa}, will 
be studied at high-$Q^2$ in two 12~GeV electron scattering experiments: E12-06-106~\cite{E12:06:106} (upcoming)
and E12-06-107~\cite{E12:06:107} (under analysis). While these experiments are complementary to our proposed CT
measurement, we point out that our measurement is sensitive to CT for baryons as well. The large number of
different final states we consider will allow us to map out the spin and isospin dependence of CT.

E12-06-107 will also measure color transparency in for protons, which complements our baryonic CT measurements.
A significant difference between this measurement and ours comes from the use of photon-induced reactions, which
will generally have larger energy transfer (and thus greater ``freezing'' of PLCs) than the majority of the 
kinematic space probed by E12-06-107.

As discussed in section~\ref{sec:ct}, the theoretical framework of PLCs suggests that both 
color transparency and the EMC effect have a common origin. Both these experiments and our 
proposed measurement will have complimentary roles in constraining that framework and in understanding
the origin of the EMC effect.

\section{Summary}

We propose a 30-day measurement using the real photon beam in Hall D, $d$, $^4$He and $^{12}$C targets, with
the GlueX detector in its standard configuration, with the goal of studying short-range correlations, transparency
and bound nucleon structure in nuclei. The use of a real photon beam as a probe provides an outstanding 
handle on reaction mechanism effects in SRC pair breakup, which complement the successful electron-scattering
SRC program in Halls A and B. We project count rates that exceed those of the 6~GeV-era SRC experiments,
allowing definitive measurements of SRC properties, the short-distance $NN$ interaction, in-medium modification,
and nuclear transparency. 

\bibliographystyle{ieeetr}
\bibliography{references.bib}

\newpage
\appendix
\section{Integrated cross-section}
\label{sec:cs}

To validate our simulation, we performed a back-of-the-envelope calculation of the rate of the 
$\gamma + n \rightarrow \pi^{-} + p$ reaction. The differential cross-section for the reaction can be approximated using the following:
\begin{equation}
\frac{d \sigma}{dt} = C \cdot f(s) \cdot f(\cos\theta_{cm}) \approx 1.25 \cdot 10^{7} \text{nb} \cdot \text{GeV}^{12} \cdot s^{-7}
 \cdot (1 - \cos \theta_{cm})^{-5} (1 + \cos \theta_{cm})^{-4},
\label{eq:cs}
\end{equation}

\noindent where the cross-section is in units of nb/GeV$^{2}$, and $\theta_{cm}$ is the polar scattering angle in the center of mass system. 

The polar scattering angle in the center of mass system $\theta_{cm}$ can be approximated as

\begin{equation}
\cos \theta_{cm} = \frac{ t - m_{\pi}^{2} + 2k_{i} \sqrt{k_{f}^{2} + m_{\pi}^{2}}}{2k_{i}k_{f}},
\label{eq:ang}
\end{equation}

\noindent where $k_{i}$, and $k_{f}$ are the center of mass momenta of incoming and outgoing particles:

\begin{equation}
k_{i} = \frac{s - m_{p}^2}{2 \sqrt{s}}
\end{equation}

\begin{equation}
k_{f} = \sqrt{\frac{(s - (m_{p}-m_{\pi})^2)\cdot(s - (m_{p}+m_{\pi})^2)} {4s}}.
\end{equation}

Equation (\ref{eq:ang}) shows that for $\theta_{cm} > 41^{\circ}$, $|t| > 2$ (GeV/c)$^{2}$. Mandelstam variables are interrelated as $s + t + u = m^2$, 
which means that for $E_{\gamma} = 9$ GeV we have $|t| > 2$ (GeV/c)$^{2}$ and $|u| > 2$ (GeV/c)$^{2}$ for almost the whole range 
of $\theta_{cm}$ between $40^{\circ}$ and $140^{\circ}$.

The cross-section (\ref{eq:cs}) in terms of $\cos\theta_{cm}$:

\begin{equation}
d \sigma = C \cdot s^{-7} \cdot f(\cos\theta_{cm}) dt = C_1 \cdot s^{-7} \cdot f(\cos\theta_{cm}) d\cos\theta_{cm},
\end{equation}

\noindent where $C_1 = 1.25 \cdot 10^{7}  \cdot 2k_{i}k_{f}$.

The total cross-section for $E_{\gamma} = 9$ GeV and $\theta_{cm}$ in the range between $40^{\circ}$ and $140^{\circ}$ is then about $2.1$ nb. The number of expected MF events per day on a carbon target is

\begin{equation}
N = \sigma_{nucl} \cdot F \cdot T \cdot t \cdot \epsilon = 880,
\end{equation}

\noindent where $F = 2 \cdot 10^{7}$ photons/s - photon flux on target, $T = 1.45 \cdot 10^{23}$ atoms/cm$^2$ - target density for $^{12}$C, $t = 24$ hours $\cdot 3600$ s/hour - time, $\epsilon = 0.64$ - detection efficiency, and $\sigma_{nucl} = \sigma \cdot A/2 \cdot A^{-1/3} = 5.5$ nb - nuclear cross-section for $^{12}$C. This is consistent within $20 \%$ with the simulation results (740 events/day for the $^{12}$C target).

\end{document}